\documentclass[lettersize,journal,comsoc]{IEEEtran}

\usepackage{tikz}
\usepackage{amsmath}

\usepackage{adjustbox}
\usepackage{booktabs}
\usepackage{balance}
\usepackage{cite}
\usepackage{dblfloatfix}
\usepackage{fontawesome5}
\usepackage{pgfplots}
\usepackage[font=small,subrefformat=simple,labelformat=simple]{subfig}
\usepackage{soul}
\usepackage{hyperref}
\usepackage[absolute,showboxes]{textpos}

\DeclareUnicodeCharacter{2212}{−}
\usepgfplotslibrary{
  groupplots,
  dateplot,
}
\usetikzlibrary{
  arrows.meta,
  colorbrewer,
  decorations.pathreplacing,
  fadings,
  fit,
  patterns,
  positioning,
}
\pgfdeclarelayer{bg}    %
\pgfdeclarelayer{fg}    %
\pgfsetlayers{bg,main,fg}  %

\hypersetup{
  colorlinks=false,
  hidelinks,
  breaklinks,
}

\graphicspath{{figs/}}

\date{}

\usepackage{pifont}
\newcommand{\cmark}{\ding{51}}%
\newcommand{\xmark}{\ding{56}}%

\usepackage{xspace}
\newcommand{\eg}{\textit{e.g.,}~}
\newcommand{\ie}{\textit{i.e.,}~}

\newcommand{\etal}{\textit{et al.}~}
\newcommand{\one}{({\em i})\xspace}
\newcommand{\two}{({\em ii})\xspace}
\newcommand{\three}{({\em iii})\xspace}

\usepackage{cleveref}
\crefname{appendix}{Appendix}{Appendices}
\crefname{equation}{Equation}{Equations}
\crefname{pluralequation}{Equations}{Equations}
\creflabelformat{equation}{#2\textup{#1}#3}
\crefname{section}{Section}{Sections}
\crefname{figure}{Figure}{Figures}
\crefname{table}{Table}{Tables}
\crefname{listing}{Listing}{Listings}

\makeatletter
 \renewcommand{\paragraph}[1]{\vspace*{0.03in}\noindent{\bf #1.}\hspace{0.25ex \@plus1ex \@minus.2ex}}
\makeatother

\makeatletter
\newcommand{\paragraphx}[1]{\vspace*{0.03in}\noindent{\bf #1}\hspace{0.25ex \@plus1ex \@minus.2ex}}
\makeatother

\makeatletter
\newcommand{\paragraphS}[1]{\vspace*{0.03in}\noindent{\bf #1?}\hspace{0.25ex \@plus1ex \@minus.2ex}}
\makeatother

\usetikzlibrary{external}
\tikzexternaldisable%
\usepackage{ifthen}

\newlength\bitboxwidth%
\newlength\bitboxheight%

\title{A Leaner and Faster Web: How CBOR Can Improve Dynamic Content Encoding in \\ JSON and DNS over HTTPS}

\author{Martine S. Lenders~\IEEEmembership{Member, IEEE},
  Carsten Bormann,
  \\
  Thomas C. Schmidt,~\IEEEmembership{Member, IEEE},
  Matthias~W\"ahlisch,~\IEEEmembership{Member, IEEE}%
\thanks{Manuscript received February 5, 2026; revised May 28, 2026; accepted Jul 31, 2026. %
This research has been supported in parts by the German Federal %
Ministry of Research, Technology and Space (BMFTR) within %
the projects PIVOT (grants 16KIS1386K and 16KIS1387), PRIMEnet (grants 16KIS1369 and 16KIS1371), and C-ray4edge (grants 16KIS1694K and 16KIS1695). %
\textit{(Corresponding Author: Martine S. Lenders.)}}%
\thanks{Martine S. Lenders is with TUD Dresden University of Technology, Dresden, Germany (email: \href{mailto:martine.lenders@tu-dresden.de}{martine.lenders@tu-dresden.de}).}%
\thanks{Carsten Bohrman is with Universit\"at Bremen TZI, Bremen, Germany (email: \href{mailto:cabo@tzi.org}{cabo@tzi.org}).}%
\thanks{Thomas C. Schmidt is with HAW Hamburg, Hamburg, Germany (email: \href{mailto:t.schmidt@haw-hamburg.de}{t.schmidt@haw-hamburg.de}).}%
\thanks{Matthias W\"ahlisch is with TUD Dresden University of Technology \& Barkhausen Institut, Dresden, Germany (email: \href{mailto:m.waehlisch@tu-dresden.de}{m.waehlisch@tu-dresden.de}).}}%

\usepackage{fancyhdr}
\fancypagestyle{GlobalFootnote}{%
  \lhead{\scriptsize \MakeUppercase{Transactions on Network and Service Management, Vol. 23, August~2026}}%
  \rhead{\scriptsize \thepage}
  \cfoot{\scriptsize This work is licensed under a Creative Commons Attribution 4.0 License. For more information, see \url{https://creativecommons.org/licenses/by/4.0/}}%
}
\begin{document}
\bstctlcite{IEEEexample:BSTcontrol}

\maketitle
\thispagestyle{empty}

\definecolor{boxgray}{rgb}{0.93,0.93,0.93}
\textblockcolor{boxgray}
\setlength{\TPboxrulesize}{0.7pt}
\setlength{\TPHorizModule}{\paperwidth}
\setlength{\TPVertModule}{\paperheight}
\TPMargin{5pt}
\begin{textblock}{0.8}(0.1,0.02)
  \noindent
  \footnotesize
  If you refer to this paper, please cite the peer-reviewed publication: Martine S. Lenders, Carsten Bormann, Thomas C. Schmidt, and Matthias~Wählisch, 2026. Leaner and Faster: The Web and DNS Can Benefit from CBOR. In \emph{IEEE Transactions on Network and Service Management (TNSM)}, Vol. 23, pp. 7904--7922, August 2026, \url{https://doi.org/10.1109/TNSM.2026.3722114}
\end{textblock}

\begin{abstract}
The Internet community has taken major efforts to decrease latency on the World Wide Web with significant improvements in accelerating content transport and in compressing static content.
Less attention, however, has been dedicated to compression of dynamic content.
Such content is commonly provided by JSON and DNS over HTTPS\@.
Dynamic content objects continue to grow in size, which increases latency and fosters the digital inequality.
In this paper, we propose to mitigate this increase by utilizing Concise Binary Object Representation~(CBOR), a standard originally designed for the constrained Internet of Things~(IoT) to restrict packet sizes and enable efficient encoding of data objects.
We provide protocol design and three new data sets for the evaluation of dynamic content, DNS, and the loading of websites.
Our key findings are the following:
\one Switching the data representation from JSON to CBOR reduces data by up to 80\%.
This size reduction can decrease loading times by up to 13.8\% when downloading large objects---even in local setups.
\two Enabling CBOR for DNS over HTTPS~(DoH) and DNS over CoAP~(DoC) reduces packet sizes significantly.
Compressing only names combined with unpacked CBOR achieves maximum gain of 52.2\%, using more complex but still lightweight Packed CBOR allows minimizing packets by up to 95.5\%.
Our lean decoder for name compression can fit into as little as 314 bytes of build~size.
Our results clearly show the potential of CBOR outside of IoT~scenarios.
Parts of this research have already influenced work within the IETF\@.
\end{abstract}


\section{Introduction}\label{sec:intro}

\IEEEPARstart{T}{he modern Web} provides access to a rich media landscape.
Video streaming services and gaming sites promise endless entertainment for everyone.
For a large portion of the global population, however, reality means struggling to open even the plainest news or utility websites.
Poor connectivity and high costs per gigabyte strike developing countries~\cite{zcpas-dwlg-14,caazr-twwwd-23,bvsz-ddcwm-25} and even rural areas in the so-called developed nations~\cite{tsfw-ebass-13}.
The latest or fastest end-devices are, likewise, inaccessible to low-income users or those of low digital affinity in all regions of the world.
Instead of investing in another device, this user group might even reuse web browsers on available devices that are not primarily intended to access the Web, such as Smart TVs or gaming consoles~\cite{e-uesh-2021}.

\begin{figure}
  \includegraphics{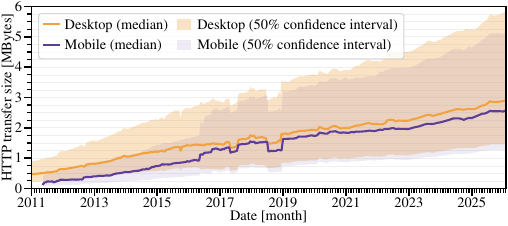}
  \centering
  \caption{
    HTTP transfer sizes of web resources when requesting a page, according to data from the HTTP Archive, November~15,~2011 to February 1, 2026~\cite{httparchive}.
  }%
  \label{fig:httparchive}
\end{figure}

Digital inequality increases because transfer sizes on the World Wide Web grew significantly during the past 30~years~\cite{vsw-fsrwh-24}.
Despite efforts to decrease traffic latencies by protocol updates such as HTTP/2~\cite{RFC-9113} or QUIC~\cite{RFC-9000}, the last decade saw an increase of HTTP transfer size per resource by almost 1.3 MBytes, according to the HTTP Archive~\cite{httparchive}, see \cref{fig:httparchive}.
While static content can easily be pre-compressed using compression techniques such as GZip or Brotli, dynamic content requires compression on the fly, leading to additional overhead and thus latency.
This dynamic content primarily stems from RESTful API requests and responses (\eg HTTP POST containing JSON and DNS over HTTPS),

Most dynamic content cannot be pre-compressed.
Neither does a web server know which data a client needs, nor can a server compress all possible content objects in advance given the sheer amount of data.
The same limitation applies to clients that compose query data based on responses they receive from a Web server.
This makes JSON and DoH messages the most obvious candidates for analyzing the usage of CBOR on the~Web.

In this paper, we suggest to complement current web standards by leaner versions---motivated by ongoing standardization activities for the constrained Internet of Things~(IoT) and bolstered by empirical measurements.
While the mainstream Web demands growing resources, the IETF has introduced protocols to support communication of constrained IoT devices~\cite{RFC-7252}.
From this IETF work, the Concise Binary Object Representation (CBOR)~\cite{RFC-8949} emerged, a new serialization format designed for constrained use cases.
CBOR employs principles known from JSON and provides structures such as arrays and maps to embed numbers, strings, booleans, or further structures.
Different from JSON, CBOR is encoded in a binary format and tailored to constrained use cases.
Consequently, CBOR aims to be implemented in limited memory space and generates only a small message size footprint.
While not being as powerful in reducing redundancies, CBOR offers also a size decrease, and is able to serialize data, including binary data, directly to the transfer medium without additional compression time (see also \cref{tab:compr-vs-cbor}).

\begin{table}
  \caption{The advantages of CBOR over compression.}%
  \label{tab:compr-vs-cbor}
  \centering
  \begin{tabular}{rcc}
    \toprule
    Feature                               & Compression           & CBOR                  \\
    \midrule
    Size decrease                         & \color{Dark2-A}\cmark & \color{Dark2-A}\cmark \\
    Binary format                         & \color{Dark2-A}\cmark & \color{Dark2-A}\cmark \\
    Sending during object serialization   & \color{Dark2-B}\xmark & \color{Dark2-A}\cmark \\
    Sending without extra computing time  & \color{Dark2-B}\xmark & \color{Dark2-A}\cmark \\
    \bottomrule
  \end{tabular}
\end{table}

\begin{figure}
  \centering
  \includegraphics{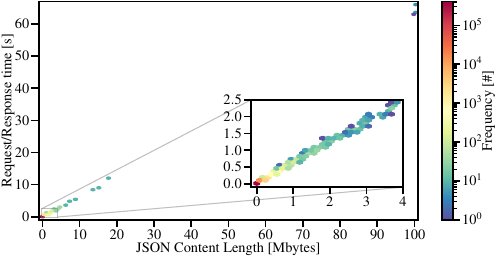}
  \caption{
    The importance of reducing sizes of JSON objects.
    The response sizes of fetching 41,269 valid JSON resources, which have been collected by the HTTP Archive and randomly selected by us. Each JSON content is fetched 10$\times$ in a local setup.
  }%
  \label{fig:resp_len-vs-plt}
\end{figure}

We address two questions:
\one Can CBOR notably improve web performance for JSON content?
\two Can CBOR optimize DNS over the Web by providing a compacter DNS message format and further schemes to decrease message sizes?
We focus on JSON and DNS because they are both relevant examples of delivering unpredictable dynamic content on the Web, which cannot be pre-compressed in a scalable way.
JSON is not the dominant content type~($\approx0.6\%$ according to HTTP Archive) in the face of video streaming but increases its share to the overall web traffic with every new front-end framework (\eg Qwik)~\cite{vgsb-cjtpc-2019,stateofjs}.
While streaming video can be avoided, the users in regions with poor Internet connectivity cannot easily avoid JSON content.
For example, news sites often side-load content in the form of large JSON objects, \eg to update their news tickers.
In \cref{fig:resp_len-vs-plt}, we illustrate the importance of reducing the size of JSON content by showing the correlation between request-response times and JSON object sizes, based on fetching locally $41{,}269$ JSON resources collected randomly from the HTTP Archive.
These JSON objects were minified, \ie stripped of all unnecessary whitespaces, but not compressed.
These request-response times add up to the page load times in real-life scenarios.

DNS over HTTPS (DoH) is another web service that delivers highly dynamic content.
Response messages are composed depending on the originating query.
Which name (and DNS records) a client requests is hard to predict.
Even if content of DNS caches is considered, prior compression is not doable because each DNS entry includes continuously changing data (the time to live), which must be preserved for consistency.
Furthermore, nearly every web access is preceded by DNS queries, accelerating the DNS resolution by reducing the amount of data will positively affect the page load times perceived by users~\cite{ck-pcdr-03,nckmw-mnpud-14,zcpas-dwlg-14,tdb-dlvitcj-23}.

To answer our research questions, we follow a measurement-driven protocol design.
We base our analysis on a large subset of JSON responses provided by the HTTP Archive, and use responses generated by the GitHub API as an example for a domain-specific conversion from JSON to CBOR\@.
To quantify the potentials of concise DNS messages, we extract DNS messages from PCAP traces of consumer-grade IoT devices~\cite{alam2019yourthings,rdcmkh2019moniotr,ppaa2020iotfinder}.
Additionally, we query names for various records via three public DNS resolvers feeding names from the Tranco list~\cite{lvtkj-tranco-2019} and use names as data from a Web PKI study~\cite{thlsw-dccdi-24} and SecSpider~\cite{ormz-qosdd-08}.

Our main contributions read.

\begin{enumerate}
\item Measurements and analysis of the web performance when CBOR would replace JSON, confirming results from previous, smaller studies. %

\item A comprehensive evaluation of our CBOR-based DNS message formats, including our new proposal of name compression combined with plain CBOR. These results influenced discussions within the IETF.

\item Three public data sets for future evaluation, and an extended taxonomy~\cite{vk-bjbss-2022} of JSON~objects to cover CBOR-based serialization and allow for coherent evaluation.

\item A runtime analysis of the Python-based parsers and serializers for both JSON and CBOR\@.
\item End-to-end measurements in distributed Web scenarios to confirm performance benefits when using CBOR\@.
\end{enumerate}

Other constrained IoT technologies, such as the Constrained Application Protocol~(CoAP)~\cite{RFC-7252} or Static Context Header Compression~(SCHC)~\cite{RFC-8724} could also provide benefits when integrated into the Web.
However, we like to offer our analysis foremost as a motivating experiment on how the efficiently encoded and largely optimized ingredients from the IoT cookbook could improve web performance.
CBOR is a good starting point to conduct such a study for the following reasons.
First, it supports extensibility without version negotiations.
This makes CBOR a drop-in replacement for JSON that can already be used today, using existing JavaScript libraries, to reduce web resource sizes and thus decrease page load times.
While CBOR has been explored in the past, we want to emphasize this in our work to encourage implementation.
Second, the binary and concise formatting of CBOR inspires deployment in further use cases, such as a DNS message format~\cite{draft-lenders-dns-cbor}.
Third, other binary formats often lack a direct mapping to JSON and a standardization in the IETF---the same body that authored the JSON specification.
Both direct JSON mapping and being a common Internet standard strongly motivate the focus on CBOR in this paper.

In the remainder of this paper, we present background on the constrained IoT with CBOR and summarize related work in \cref{sec:background}.
 In \cref{sec:eval-j2c}, we evaluate the potential of CBOR as a drop-in replacement for JSON, as well as how features of CBOR can improve performance even further.
 We investigate the compact CBOR DNS message format and our name compression in \cref{sec:eval-c4d}.
We report on the implementation along with a comparative evaluation in \cref{sec:impl}.
We validate and connect all our findings in real-world Web scenarios in \cref{sec:eval-e2e}, and discuss our results in the context of further optimization potentials with CBOR for both DNS and the Web in \cref{sec:discussion}.
Last, we conclude with an outlook on future work in \cref{sec:conclusion}.

\section{Background and Related Work}\label{sec:background}

\noindent In this section, we give a brief overview of the constrained Internet of Things (IoT) and the technologies standardized within the IETF\@.
Then we summarize related work.

\subsection{The Constrained IoT}\label{sec:background:bg}
\paragraph{An overview of the constrained IoT}
Over the last two decades, the IETF and other standardization bodies enabled Internet communication on tiny devices.
These devices face multiple constraints as they are often powered by batteries, \ie are low-power, have low memory in the kBytes range, and are typically connected via low-power lossy links such as IEEE~802.15.4 or LoRaWAN~\cite{RFC-7228}.
At the low-end, adaption layers such as 6LoWPAN~\cite{RFC-4944} or Static Context Header Compression and Fragmentation (SCHC)~\cite{RFC-8724} allow for sending full IPv6 packets over links of $\approx 100$ bytes maximum frame size.
The Constrained Application Protocol~(CoAP)~\cite{RFC-7252} transfers RESTful web applications from the heavy-weight HTTP while providing proxy capabilities for compatibility and integration with HTTP, \eg via WebSockets~\cite{RFC-8323}.

\begin{figure}
  \centering%
  \subfloat[An Integer.]{%
    \centering%
    \includegraphics{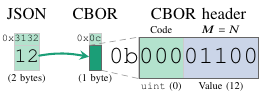}\label{fig:cbor-example:sint}
  }\\
  \subfloat[A String.]{%
    \centering%
    \includegraphics{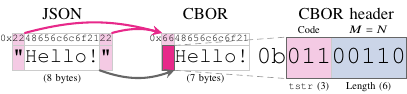}\label{fig:cbor-example:str}
  }
  \caption{Examples for transforming JSON elements to CBOR\@.}%
  \label{fig:cbor-example}%
\end{figure}

\paragraphx{Constrained Binary Object Representation (CBOR)}~\cite{RFC-8949} encodes application data for the constrained IoT in an efficient, non-inflating way.
Its binary encoding allows for a more concise representation of application data compared to, \eg JSON~\cite{RFC-8259}, while providing the same format semantics as JSON\@.
Some example conversions from JSON elements can be seen in \cref{fig:cbor-example}.
CBOR encodes elements in binary using 8 types identified by a 3-bit numerical code: \emph{unsigned integers} (0), \emph{negative integers} (1), \emph{byte strings} (2), \emph{text strings} (3), \emph{arrays} (4), \emph{maps} (5), \emph{tags} (6), as well as \emph{simple values and floats} (7).
Following the \emph{code} for the major type a 5-bit field $M$ which we call the \emph{indicator} describes the length of the argument $N$.
If $M \leq 23$, then $N = M$, otherwise if $M$ is 24, 25, 26, 27, $N$ is encoded in the next 1, 2, 4, or 8 bytes, respectively, in network byte order.
The meaning of $N$ depends on the major type and on $M$.
Typically, for primitive-typed CBOR elements, $N$ expresses the value itself but for structural CBOR elements or strings it denotes the length of the structure or string.
Simple values are values of major type 7 with $N \leq 255$.
Some of them are used for booleans or other literals such as \emph{null}, but most are an additional number range to unsigned integers.
For further details, please see Appendix~\ref{sec:cbor-semantics}.
Tags extend this functionality and are used as functions or markers that are represented as a numerical value prepending other CBOR elements.
For instance, tag 24 marks embedded CBOR binaries (\eg \texttt{24(h'426567')} marks the CBOR byte string ``eg'') or tag 34 marks text strings as a base64-encoded binary (\eg \texttt{34("ZS5nLg==")} marks the base64-encoded byte string ``eg'')~\cite{RFC-8949}.  %
While the transport format for CBOR is binary, a textual diagnostic notation for CBOR (used in the examples above) serves debugging and documentation.
Packed CBOR, among other things, provides a mechanism to further reduce redundancies with in a CBOR-encoded object~\cite{draft-ietf-cbor-packed}.
It prepends one or more \emph{packing tables} within a \emph{table setup tag}, containing the parameters for compression, in form of CBOR arrays.
The remainder of the object, the \emph{rump}, then references the indices of those packing tables using simple values (for value references) or tag 6 (both value references, but also, \eg prefix and suffix concatenation).

\paragraphS{What can the World Wide Web learn from the constrained IoT}
The constrained IoT aims to decrease the size of packets.
Every byte, sometimes even bit, is accounted for.
Much of the Internet usage already ossified to the use of HTTP, so a transition, \eg to CoAP is not to be expected.
However, we envision notable benefits in reducing HTTP transfer size with CBOR where JSON, or even XML, is used for dynamic content today.
We also envision acceleration of the Web by encoding DNS messages in CBOR\@.
Compared to JSON, it is \one a binary format and thus does not need special transport encodings for byte data, such as base64, \two its integer encodings are generally more compact, and \three mostly uses only one control token at the beginning of each element for its structure, which can decrease both content length and parsing complexity.
CBOR was also chosen for the European Health Certificate Specification, the specification that defined the European COVID-19 vaccination certificates~\cite{eu-dcc-21} and is used by the ISO/IEC 18013-5 standard for mobile \mbox{driving~licenses}~\cite{iso-18013-5}. %

\subsection{Related Work}\label{sec:related-work}

\paragraph{Optimizing the World Wide Web}
Efforts to optimize the World Wide Web are nearly as old as the Web itself. In 1997, Mogul \etal~\cite{mdfk-pbdedch-97} proposed to reduce HTTP transfer sizes by refreshing cached HTTP responses delta-compressed.
DEFLATE-based compression~\cite{RFC-1951} is another scheme to reduce HTTP transfer sizes.
This, however, introduces more complexity and thus may introduce security problems~\cite{pbws-chnss-2015}.
DEFLATE, GZip, or Brotli, or even more edge-focused approaches, such as Co4U~\cite{zczxw-cerhm-24}, cannot be streamed in-line by constructing an object while sending it.
Naylor \etal~\cite{nflgm-csh-2014} found that the advent of HTTPS correlated with a reduction of in-network caching and compression.
Optimizations at the protocol layer were started more than ten years ago.
First, SPDY introduced---among others---multiplexing of HTTP messages in one pass and reduced header sizes for HTTP with a binary encoding~\cite{draft-mbelshe-httpbis-spdy,wbkw-hss-14}, which led to HTTP/2~\cite{RFC-9113}.
Thereafter, transport of the Web moved toward QUIC~\cite{RFC-9000} and ultimately to HTTP/3~\cite{RFC-9114}.
Wolsing \etal~\cite{wrwh-ppwop-19} confirmed QUIC performance by comparing it to highly optimized TCP stacks.
Sengupta \etal~\cite{skffb-ciqed-24} specifically looked at the effects on page load times from QUIC-induced interactions of HTTP/3 preceded by DNS over QUIC (DoQ).
In this paper, we focus on the potentials of IETF IoT~protocols.
To that end, G\"{u}ndo\u{g}an \etal~\cite{gasw-cosit-22} propose a framework to provide Information Centric Networking (ICN) for IP networks using the IoT application layer protocols CoAP and OSCORE\@.
While our work also proposes to use IoT technologies, we keep the web-related protocols intact and only change the content encoding at the application layer.
Sharma \etal~\cite{sp-cmsct-25} propose IoT-C$^2$, a framework to cache dynamic IoT content objects at Edge servers to reduce latencies.
However, the nature of dynamic content still requires regular cache refreshes.
Using our proposal reduces latencies for cache refreshes and even device-to-edge latencies, decreasing the overall latency even more.
At the front-end side, frameworks such as \emph{Accelerated Mobile Pages} (AMP)~\cite{openjs-amp} improve the performance for mobile web browsing.
As frameworks such as these typically use JSON to serialize their messages, CBOR---which includes the semantics of JSON---provides a good drop-in replacement~for~them.

\paragraph{Digital inequality and web page reduction}
The modern Web fosters digital inequality~\cite{pavsz-tfwdr-24,bvsz-ddcwm-25}.
Outdated devices may still be more powerful than constrained IoT~devices~\cite{RFC-7228} but have less hardware resources than recent devices and thus struggle with common web pages.
Political efforts to decrease disparity might also face difficulties~\cite{mskzz-ecafa-2024}, though proposals exist to bridge the gap with wireless 5GHz and mmWave broadband~\cite{gcm-ngwbd-25}.
A common proposal is to reduce sizes by either cleaning up the delivered JavaScript~\cite{caazr-twwwd-23,chhir-bnbam-23} or by re-thinking HTML web page construction in general.
For example, the \emph{smolweb} movement aims to de-clutter HTML from CSS, JavaScript, and non-semantic tags~\cite{smolweb}.
Even alternative formats to HTML, such as MAML, have been proposed~\cite{pavsz-tfwdr-24}.
Our work supports sustainable, inclusive web communication since we aim to reduce load by protocol design, making the Web more~accessible.
Recently, the \emph{green} working group~\cite{green-wg} within the IETF and the \emph{sustain} research group~\cite{sustain-rg} within the IRTF picked up work for a more sustainable and energy-efficient Internet including and beyond the protocol layers.

\paragraph{Comparing JSON to concise binary formats}
Iglesias-Urkia \etal~\cite{icmbu-ieswi-19} showed in a Smart Grid IEC~61850 context that supplanting HTTP and JSON with CoAP and CBOR is feasible and can decrease network overhead as long as messages stay small enough.
Viotti and Kinderkhedia~\cite{vk-bjbss-2022,vk-sjbss-2022} compared CBOR against JSON and other binary formats, including DEFLATE-based compression.
They concluded that for very redundant data, CBOR is not much smaller than JSON when compressed with DEFLATE\@.
If CBOR, however, is not redundant it can be much smaller than zipped JSON\@.
The selection of objects used in this study, though, was quite limited and did not explore all the advantages of CBOR over JSON\@.
We base our classification of JSON objects on their proposed taxonomy~\cite{vk-bjbss-2022}, and provide a comprehensive analysis.
Gudia and Singh~\cite{gs-ccejd-2024} proposed an encoding for JSON-LD in CBOR that provides savings of up to $95.1\%$ of network overhead.

\paragraph{Concise DNS message formats}
Dikshit \etal~\cite{dkfsb-edrrt-24} recently noted, that DNS messages are growing in size but only analyzed DNS over TCP as a potential fallback solution.
DNS poses consequently an additional latency bottleneck for the World Wide Web~\cite{ck-pcdr-03,nckmw-mnpud-14,zcpas-dwlg-14,tdb-dlvitcj-23}.
With the advent of DNS over HTTPS (DoH)~\cite{bcaft-escdoh-19,RFC-8484} DNS also was integrated into the Web.
As such, reducing DNS message sizes to decrease latency is desirable.
DNS already comes with a compression mechanism that references name suffixes within the message~\cite{RFC-1035}.
A brief overview over the DNS message format and name compression can be found in Appendix~\ref{sec:dns-msg}.
Klauck and Kirsche~\cite{kk-ednsmc-23} proposed to stick with the classic DNS format and to rather extend name compression for other redundant fields, such as TTL and record types.
However, since the proposal lacks type markers it remains restricted to the specific use case of DNS Service Discovery (DNS-SD) over 6LoWPAN\@.
With DNS over CoAP (DoC)~\cite{lagns-snrid-23,RFC-9953}, encrypted and RESTful DNS also entered the constrained IoT domain.
For this, we propose a new DNS message format based on CBOR~\cite{draft-lenders-dns-cbor}.
Note, that this format differs from the CBOR-based C-DNS, specified in RFC~8618~\cite{RFC-8618}.
C-DNS uses a globally applied compression mechanism which is designed to store and exchange DNS packet \emph{captures} in CBOR, but not to exchange single DNS messages over the network.
Lemogue \etal~\cite{lmtb-tfcdmcs-23} proposed an alternative to our CBOR draft, which includes name compression.
However, their name compression proposal is more similar to Packed CBOR~\cite{draft-ietf-cbor-packed} and introduces a dedicated structure.
Furthermore, it introduces unnecessary overhead, as it uses maps over arrays to implement the suffix table, with the reference key just being consecutive integers carried explicitly in the message, rather than using implicit array indices.

\begin{figure*}
  \vskip-0.5em
  \centering
  \setlength{\belowcaptionskip}{-1.0em}
  \includegraphics{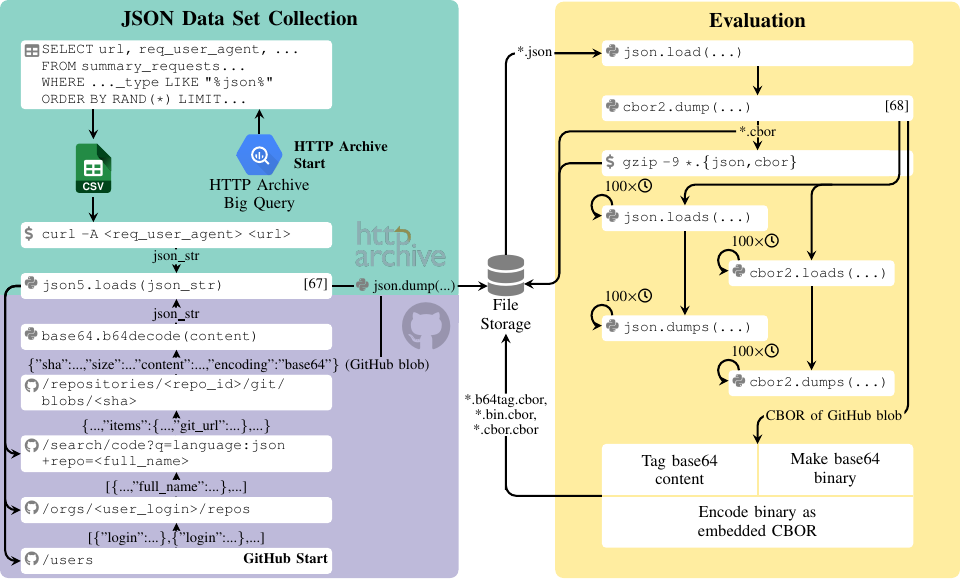}
  \caption{Our JSON data set collection and measurement method.}%
  \label{fig:json-collect}
\end{figure*}

\pagebreak
\section{Evaluating the Use of \\ CBOR for Object Encoding}\label{sec:eval-j2c}

\noindent We now compare CBOR with JSON and explore the potential of CBOR when using its tags and byte strings.
CBOR was developed as a concise, binary alternative to JSON, which is typically used to encode \emph{dynamic} web objects---\ie objects that cannot be generated beforehand---as, \eg used in REST API calls.
Vargas \etal~\cite{vgsb-cjtpc-2019} measured a steadily growing JSON to HTML request ratio for HTTP responses at a CDN in 2019. %
We were able to confirm this growth in the years since with the HTTP Archive Big Query data set~\cite{httparchive-bigquery}.
We gather generic JSONs for that data set.
After an exploration of our collected data set, we compare the request/response times of CBOR-encoded objects with those times we gathered for JSON-encoded objects in \cref{fig:resp_len-vs-plt}.
We then explore the absolute bytes that can be saved using CBOR over JSON, and the relative gain.
We also convert domain-specific JSONs, based on the GitHub REST API, to evaluate the potential of CBOR features such as byte strings and tags.

\subsection{Data Set Collection and Method}\label{sec:eval-j2c:method}

\noindent To evaluate the potential for CBOR encoding over the Web, we randomly sampled the September 2024 version of the HTTP Archive data set~\cite{httparchive} for responses that have the content type containing the string ``json'', \ie not just \texttt{application/json}.
For that we used Google Big Query~\cite{httparchive-bigquery}, the SQL-based interface to access the structured HTTP Archive data.
We sampled uniformly distributed from all rows of the ``desktop''-version of the \texttt{requests} table for September 2024\footnote{The schema of this database changed during our study. This information now resides in the \texttt{crawl} table, but the information is still there.}.
The sampling was done to remain within the monthly 1~terabyte quota of Big Query.
Using curl v7.68~\cite{curl} we downloaded these responses in November 2024, using the URL and user agent identification from the HTTP Archive entries and tried to parse them using \texttt{json5} v0.9.24~\cite{pyjson5}, which is an exceptionally robust and liberal parser (it understands non-standard JSON, \eg with comments).
The randomized sampling and the usage of the more liberal \texttt{json5} parser reduces a bias to certain types of objects, even though we initially filtered for content types to limit the initial pool of negatives.
Nevertheless, the data set may not be perfectly representative sample of web JSON traffic.
The detailed collection and measurement methods are shown in \cref{fig:json-collect}.

If we can successfully parse the JSON response, we store the parsed Python objects as a uniquely identified JSON file.
We discard empty responses or responses that are not parsable as JSON\@.
This still leaves some files that were interpreted as plain JSON strings by the liberal parser even though they were other formats, \eg default error messages by the HTTP server in HTML\@.
As such, we also exclude response that the parser identified as flat string.

We re-code the gathered JSON files for processing our evaluation.
We use the Python \texttt{json} library and the \texttt{cbor2} library v5.6.2~\cite{cbor2} (Python v3.11.2).
To encode the Python objects to minified JSON, we use the \texttt{json.dumps()} method of Python because it is faster, and we do not have any uncommon formats anymore. %

Additionally, we randomly sampled the GitHub API between January 15, 2024, and February 15, 2024---accounting for the rate limiting of the GitHub API---for the \emph{base64}-encoded Git blob representation of JSON files that are publicly version-controlled on GitHub.
These API objects provide a very illustrative example to investigate the potential of CBOR features such as tags and byte strings for REST APIs in our analysis.
We successively transform the CBOR object of this \emph{base64}-encoded representation to an increasingly concise representation.
\one We use tag 34 to mark the content as \emph{base64} and remove the now redundant \texttt{encoding} property.
\two We remove tag 34 again, \emph{base64}-decode the content, store it as a byte string, and remove the \texttt{size} property as it is now also encoded in the $N$ of the byte string (see \cref{sec:background}).
\three If available, we replace the JSON in the content with the CBOR, and mark it with tag 24 as an embedded CBOR data item.
This last step investigates an alternative history in which JSON files stored on GitHub were encoded as CBOR files in the first~place.

\begin{figure}
  \includegraphics{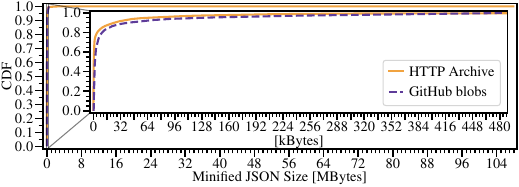}
  \caption{The overall size distribution of JSON objects stripped of all unnecessary whitespaces (minified JSON) in our data sets in bytes.}%
  \label{fig:json_sizes}
\end{figure}

\begin{table}
  \centering
  \caption{Taxonomy of our collected JSON HTTP Archive data set according to Viotti and Kinderkhedia~\cite{vk-bjbss-2022}.}%
  \label{fig:json_taxonomy}
  \subfloat[Size Tier]{%
  \centering
\begin{tabular}{rr}
\toprule
Tier & Ratio \\
\midrule
1 ($< 100$ bytes) & $32.5\%$ \\
2 ($< 1000$ bytes) & $34.4\%$ \\
3 ($\geq 1000$ bytes) & $33.1\%$ \\
\bottomrule
\end{tabular}\label{fig:json_taxonomy_tier}%
}\hspace{1em}%
  \subfloat[Content Type]{%
  \centering
\begin{tabular}{rr}
\toprule
Content Type & Ratio \\
\midrule
Numeric & $17.1\%$ \\
Structural & $5.7\%$ \\
Textual & $77.2\%$ \\
\bottomrule
\end{tabular}\label{fig:json_taxonomy_content_type}%
  }\\
  \subfloat[Redundancy]{%
  \centering
\begin{tabular}{rr}
\toprule
Redundancy & Ratio \\
\midrule
Non-redundant & $64.6\%$ \\
Redundant & $35.4\%$ \\
\bottomrule
\end{tabular}\label{fig:json_taxonomy_redundancy}%
}\hspace{1em}%
  \subfloat[Structure]{%
  \centering
\begin{tabular}{rr}
\toprule
Structure & Ratio \\
\midrule
Flat & $73.6\%$ \\
Nested & $26.4\%$ \\
\bottomrule
\end{tabular}\label{fig:json_taxonomy_structure}%
  }
\end{table}

\subsection{Data Corpus}\label{sec:eval-j2c:expl}
\noindent Our data set comprises $120{,}929$ JSON responses collected from the HTTP Archive and $128{,}643$ GitHub blob objects.
The size distributions for both data sets can be seen in \cref{fig:json_sizes}.
While the HTTP Archive data set contains overall larger objects of up to $107.8$~MBytes, both data sets follow a similar distribution of sizes.
The GitHub blob data set maximum of $482.5$~kBytes is near the 99.6\% percentiles for the HTTP Archive data set of $482.1$~kBytes.

We categorize the more diverse HTTP Archive data according to the taxonomy for JSON objects proposed by
Viotti and Kinderkhedia~\cite{vk-bjbss-2022} for benchmarking JSON-compatible binary serialization formats.
Four categories exist in this taxonomy:
\textbf{Tier}, \textbf{content type}, \textbf{redundancy}, and \textbf{structure}.
The \textbf{tier} describes the byte size of the JSON file after minification.
\emph{Tier 1} are JSON files $< 100~\text{bytes}$, $100 \le $ \emph{tier 2} $< 1{,}000~\text{bytes}$, and \emph{tier 3} $\ge 1{,}000~\text{bytes}$.
The \textbf{content type} describes the most prevalent type in the object, \ie string (\emph{textual}), number (\emph{numeric}), or boolean or \texttt{null} (\emph{boolean}), or arrays/objects (\emph{structural}).
\textbf{Redundancy} describes the repetitiveness of values in an object as \emph{redundant} or \emph{non-redundant}.
\textbf{Structure} describes whether a JSON object is \emph{nested} or \emph{flat}.

The relative distribution of the different categories available in our data set is shown in \cref{fig:json_taxonomy}.
The data set consists largely of JSON files exhibiting flat, non-redundant, textual content.
$50{,}770$ files (42.0\% of all files) fall in these categories.
Size-wise, we see an equal distribution.
This is particularly interesting as the typical threshold where compression such as GZip or Brotli creates gain is about 1 kBytes---compression is effectively applicable to only one third of our \emph{randomly sampled} data.
The high share of non-redundant JSONs decreases the potentials for compression further.
Given the high coverage of the HTTP Archive, this further confirms that dynamically generated JSON files are rarely suitable for classic compression.

\subsection{Serialization and Parsing Runtime}\label{sec:eval-j2c:runtime}
\begin{figure}
  \centering
  \subfloat[Serialization.]{\includegraphics{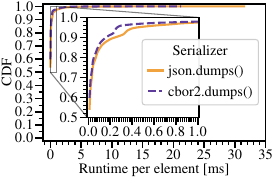}\label{fig:json2cbor_times:dumps}}%
  \subfloat[Parsing.]{\includegraphics{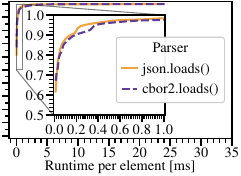}\label{fig:json2cbor_times:loads}}
  \caption{Average runtime per element over 100 runs in milliseconds ($\text{ms}$) for serialization (``\texttt{dumps()}'') and parsing (``\texttt{loads()}'') a Python dictionary with \texttt{json}/\texttt{cbor2}. Elements are each dictionary, list, CBOR tag, or primitive value within the dictionary.}%
  \label{fig:json2cbor_times}
\end{figure}

\noindent We run the \texttt{dumps()} (serialization) and \texttt{loads()} (parsing) functions of the \texttt{json} and \texttt{cbor} libraries 100 times each on a 1.2 GHz ARM Cortex-A53-based quad-core machine over the objects of the GitHub data set. %
The relatively slow platform was chosen to get a significant spread of runtimes for better comparability.
\cref{fig:json2cbor_times} plots the average runtime per JSON element.
Both libraries perform very similar, with mean runtimes being $0.12 \pm 0.63$ ms for JSON encoding and $0.07 \pm 0.33$ ms for JSON decoding.
For CBOR we measure $0.09 \pm 0.44$ ms for encoding and $0.11 \pm 0.47$ ms for decoding.
For comparison, we repeat this evaluation on a modern server CPU, an AMD EPYC 7702 (not shown), which runs an order of magnitude faster but verifies the relative results on the Cortex-A53.
This confirms that changing the JSON serializer for a CBOR one, even with relatively slow implementations, should not have any impact on the delay.

\subsection{Compression Gain and Byte Savings}\label{sec:eval-j2c:gain-savings}

\noindent As shown in \cref{fig:resp_len-vs-plt}, the request/response advantage is directly linked to the size of a response.
To determine size advantages, we use the byte savings $b$ as absolute metric and gain $g$ as a relative metric that is derived from $b$, where
``$\text{Minified JSON size}$'' is the size of the minified JSON file in bytes and ``$\text{CBOR size}$'' the size of the CBOR file in bytes.

\begin{align}
  b &= \text{Minified JSON size} - \text{CBOR size} \label{eq:savings}
\end{align}
\begin{align}
  g &= \frac{b}{\text{Minified JSON size}} \label{eq:gain}
\end{align}

A common way to grasp $g$ is to understand it as the relative amount of the original size that now could be used for more data, \ie a gain of 100\% means, the whole original message fits now into the new format,
50\% provides space for half the original message, \emph{etc}.
A negative gain in turn means, that an additional inverse of the percentage of the message size is required for the compressed message, \ie a gain of -50\% means that the ``compressed'' message needs 50\% more space than the original message.

\begin{figure}
  \centering
  \subfloat[Byte savings $b$.]{\includegraphics{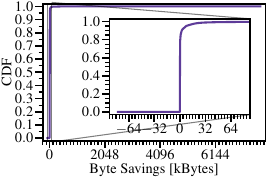}\label{fig:json2cbor-savings}}%
  \hspace{1mm}%
  \subfloat[Gain $g$.]{\includegraphics{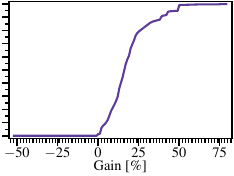}\label{fig:json2cbor-gain}}\hspace{1mm}%
  \caption{Advantages of CBOR over JSON for all the sampled HTTP Archive JSONs.%
  }%
  \label{fig:json2cbor-comp}
\end{figure}

\paragraph{Benefits of simple conversion}
\Cref{fig:json2cbor-comp} shows byte savings $b$ and gain $g$ for the JSON files from the HTTP Archive.
CBOR saves on average $15.2 \pm 405.5$~kBytes ($18.8 \pm 12.8\%$ gain), with a maximum of $7.6$~MBytes ($80.0\%$ gain).

Encoding in CBOR is not always beneficial.
Considering the different taxonomies, the JSON file with the highest inflation ($-77.6$ kBytes saving) is \emph{numeric}, \emph{nested}, and \emph{redundant}, meaning it contains a high number of repeating floating point numbers in arrays or maps.
Our CBOR encoder encodes all floats to 8 byte double precision floats, \ie a JSON-encoded ``5.5'' (3 bytes) encodes to CBOR as \texttt{fb~4016000000000000} (9 bytes).
Ideal encoding would make this number as half precision float \texttt{f9~4580} (3 bytes).
This also explains our minimum of gain $-52.9\%$ in \cref{fig:json2cbor-gain}.
We confirmed this result with a more deterministic but slower CBOR encoder that fits the floating point numbers in size (not shown):
A small but not very significant advantage was visible.
Binary float encoding can, on the other hand, also be beneficial, \eg an ASCII ``0.000005428494284942873'' (23~bytes) or ``5.428494284942873e-06'' (21 bytes) encode to \texttt{fb~3ed6c4cd259b6807} (9 bytes), even when using double precision floats.

\begin{figure}
  \centering
  \subfloat[Byte savings $b$.]{\includegraphics{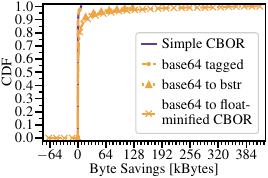}\label{fig:json2cbor-blobs-savings}}
  \hspace{1mm}%
  \subfloat[Gain $g$.]{\includegraphics{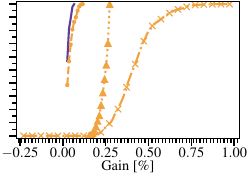}\label{fig:json2cbor-blobs-gain}}\hspace{1mm}%
  \caption{Advantages of CBOR over JSON for GitHub binary large object (blob) representations. Several encoding schemes are displayed for the binary part of the blob.}%
  \label{fig:json2cbor-blobs-comp}
\end{figure}

\paragraph{Incremental benefits of domain-specific conversion}
To analyze the impact from tags and byte strings (see \cref{sec:eval-j2c:method}), gain and byte savings for our domain-specific GitHub API data set are displayed in \cref{fig:json2cbor-blobs-comp}.
When adding a tag to mark the \textit{base64} text string (\emph{base64 tagged}), we only see small advantages with a maximum gain of $4.7\%$  ($14$ bytes)  over encoding directly to CBOR (\emph{simple CBOR}).
Removing the property subtracts $16$ bytes, the tag then adds $2$ bytes again, which sums up to saving $14$ bytes.
If we decode the \textit{base64} text string to a byte string (\emph{base64 to bstr}), we see a significant increase with a mean gain of  $24.6\%$ and  a maximum of $27.4\%$  ($6.0$ kBytes mean,  $132.3$ kBytes max).
Replacing the embedded JSON with CBOR using the smallest possible float size (\emph{base64-JSON to float-minified CBOR}) increases the space efficiency even more with a mean gain of $42.0\%$ and a maximum of $98.8\%$ ($11.8$ kBytes mean, $414.7$ kBytes max).

\subsection{Request/Response Times}\label{sec:eval-j2c:times}

\begin{figure}
  \centering
  \includegraphics{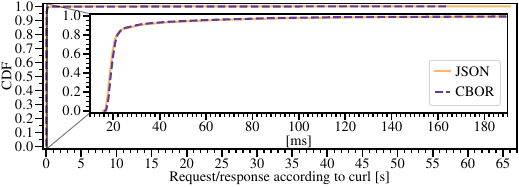}
  \caption{CDF for request/response times from a local Web server using curl.}%
  \label{fig:j2c-page-load-times}
\end{figure}

\begin{figure*}
  \centering
  \includegraphics{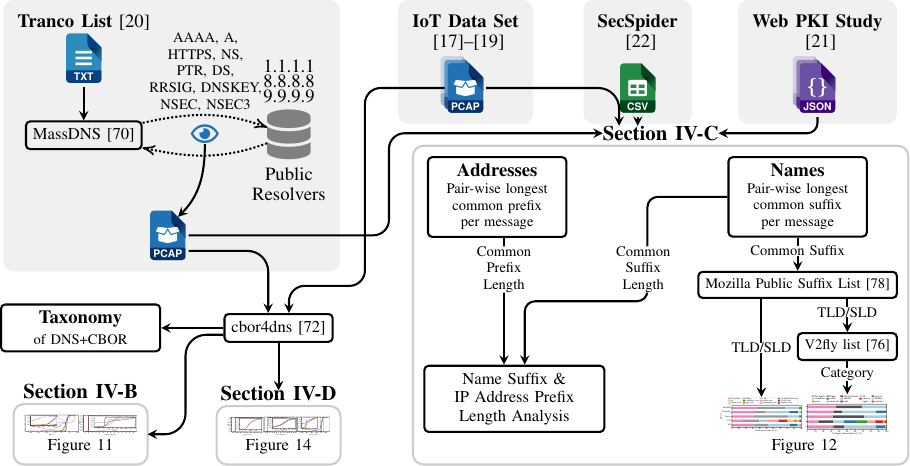}
  \caption{Our method to evaluate the \texttt{application/dns+cbor} media type.}%
  \label{fig:dns-collect}
\end{figure*}

\noindent We now look into how the request/response times in a local setup improve when objects are encoded in CBOR\@.
The local setup was chosen, as a complete CBOR front-end for browsers is out-of-scope of this paper.
We use a uniformly distributed sample of 41,269 JSON objects of our HTTP Archive data set.
This downsizing happened primarily to reduce experiment runtime.
From a local web server, we serve the uncompressed minified JSON or converted CBOR from our evaluation in \cref{sec:eval-j2c:gain-savings}, depending on experiment parametrization.
The path of the objects corresponds to the URL from which we originally downloaded the JSON\@.
We request each of these original URLs $10\times$ to decrease the impact of measurement artifacts and measure the request/response times.
We use the standard command-line tool curl~\cite{curl} to measure these request/response times.
The URLs are redirected to our local web server using the \verb$--connect-to$ parameter of curl.
The metric used as request/response time is the \texttt{time\_total} variable of curl.

We show the request/response times in \cref{fig:j2c-page-load-times}.
In the maximum we can save up to $66.0~\text{s}-56.9~\text{s} = 9.1~\text{s}$, a reduction by $13.8\%$, in this \emph{local} setup and still $187.0~\text{ms}-173.5~\text{ms}=13.5~\text{ms}$ at the 99\% percentile when using CBOR instead of JSON
The maximum stems from the same object, which consists of $100.0$~MBytes in minified JSON which was served from the CDN of a US SaaS company in September 2024.  %
The objects around the 99\% percentile are $\approx 400$~kBytes.
These measurements also served as the basis of our illustrative introduction example in \cref{fig:resp_len-vs-plt}.

\section{Evaluating CBOR as DNS Message Format}\label{sec:eval-c4d}
\noindent In this section, we analyze the performance of our proposal within the IETF for a CBOR~format for DNS~messages~\cite{draft-lenders-dns-cbor}.
Since web traffic usually triggers multiple DNS~requests, we argue that comprehensive, domain-specific CBOR~support could introduce benefits for the whole system.
After the initial data set exploration, we compare gain and byte savings of this format, evaluate the potentials of not carrying names and IP addresses explicitly, and offer an improvement to the unpacked format, based on this evaluation.
This work directly informed our IETF standardization.
We adopted one of our proposals and continuously improved upon it from version 9 of our draft~onward~\cite{draft-lenders-dns-cbor}.

\begin{figure*}
  \centering
  \adjustbox{clip=True, trim=0 9.8em 0 0}{\includegraphics{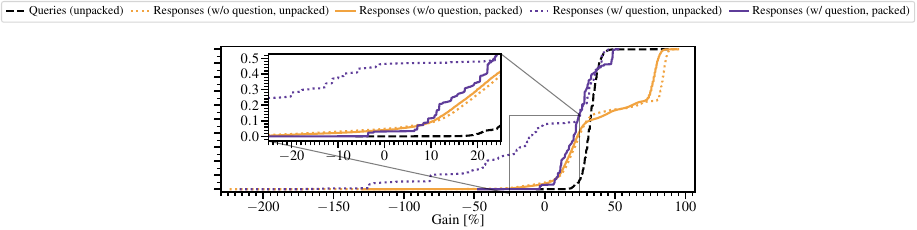}}\\[-1em]
  \subfloat[Byte savings $b$.]{\includegraphics{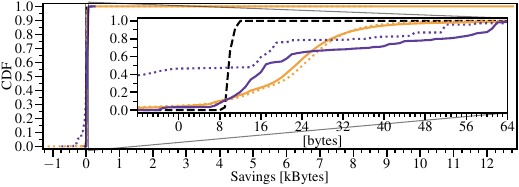}\label{fig:eval-c4d:savings}}\hspace{1mm}%
  \subfloat[Gain $g$.]{\includegraphics{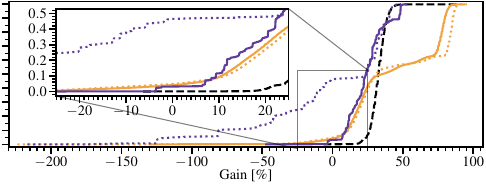}\label{fig:eval-c4d:gain}}\hspace{1mm}
  \caption{Advantages of \texttt{application/dns+cbor} over the classic DNS message format (\texttt{application/dns-message}). If the response elides the question section, it is in the ``w/o question'' group, otherwise it is in the ``w/ question'' group. ``Packed'' denotes responses encoded with the \texttt{application/dns+cbor;packed}, otherwise they are encoded with the plain \texttt{application/dns+cbor}.}%
  \label{fig:eval-c4d}
\end{figure*}

\subsection{Data Set Collection and Method}
\noindent Our evaluation is based on 4 data sets, see \cref{fig:dns-collect}.

\paragraph{IoT traffic dumps}
This data set was used in our prior work~\cite{lagns-snrid-23} to evaluate DNS over CoAP~\cite{RFC-9953} and to classify IoT domain names~\cite{alabk-tbuid-25}.
It comprises packet captures from three studies in physical IoT testbeds, Yourthings~\cite{alam2019yourthings}, IoTFinder~\cite{ppaa2020iotfinder}, and MonIoTr~\cite{rdcmkh2019moniotr}, each includes traffic from real consumer-grade IoT~devices.
In contrast to the other 3 data sets, these data sets also contain multicast DNS~(mDNS) traces from DNS Service Discovery~(DNS-SD).

\paragraph{Tranco list}
We query AAAA, A, HTTPS, NS, PTR, DS, RRSIG, DNSKEY, NSEC, and NSEC3 records for each name in the full Tranco list from December 14, 2023~\cite{lvtkj-tranco-2019} %
via three public resolvers using MassDNS~\cite{massdns} from a vantage point in Hamburg, Germany.
If we encounter a CNAME record, we also query those records for that name.
The selection of DNS record types reflects commonly queried records in the larger Internet~\cite{lagns-snrid-23} and most of the DNSSEC record types.
We record the DNS traffic as PCAPs to allow for a detailed analysis.

\paragraph{SecSpider}
This data set includes data from SecSpider~\cite{ormz-qosdd-08} gathered on December 12, 2023.
We pair up names, record types, and record data of all SOA, NS, TLSA, DS, and RRSIG records.
We use this data set only for the analysis of name compression because it does not contain full DNS messages and there were no IP addresses to pair up.

\paragraph{Web PKI study}
This data set includes DNS queries and answers for A, CAA, TLSA, and SOA records based on all entries in the Tranco list, collected in mid 2023.
The data was provided by Tehrani~\etal~\cite{thlsw-dccdi-24}.
We use this data set only for the name compression analysis.

We include the IoT traffic dumps to consider the original constrained use case.
Even though these devices are not constrained, their behavior, and thus also their name resolution, reflects typical machine-to-machine communication patterns.
The other sets serve to expand the data set for the proposed Web use case.

We verified our analysis using the DNS-OARC ``Day in the Live'' (DITL) traces of the DNS root servers collected between April 9 and 11, 2024~\cite{oarc-ditl}.
We found comparable results to the evaluation in this section, so we deem our results representative for typical Internet DNS traffic.
To ensure reproducibility, however, we did not include this undisclosed data set.

\paragraph{Setup}
We summarize our setup in \cref{fig:dns-collect}.
In total, there are over 454 million DNS queries and 68 million DNS responses in the PCAP traces in \emph{Tranco} and \emph{IoT}, with over 453 million queries and 64 million responses coming from \emph{Tranco}.
3\% of these 68 million responses cannot be mapped to a query.
These stem from the IoTFinder part of \emph{IoT} which contains only DNS responses~\cite{ppaa2020iotfinder}, but there were also $\approx$17600 previously mapped retransmission of responses in \emph{Tranco}.
We convert DNS over UDP packets, mDNS packets, and DNS over TCP streams using our \texttt{cbor4dns} tool~\cite{cbor4dns}, which was based on version 6 of the draft~\cite{draft-lenders-dns-cbor} at the time of conversion.

To answer the question of how much of the name suffixes and IP prefixes are redundant in typical DNS messages, we extend our taxonomy from \cref{sec:eval-j2c:expl} to consider special features of CBOR\@. %
Finally, we apply the lessons learned from our investigation and analyze two name compression schemes for unpacked \texttt{application/dns+cbor} and use modified versions of \texttt{cbor4dns} to analyze these schemes.

\subsection{The application/dns+cbor DNS Message Format}\label{sec:eval-c4d:dns+cbor}

\noindent The use case of the CBOR DNS message format~\cite{draft-lenders-dns-cbor} is to make DNS messages more concise for constrained networks.
For sake of conciseness it differs from JSON-based transport formats, such as the one proposed in RFC~8427~\cite{RFC-8427}, the JSON API for DoH by Google~\cite{google-dns-json-api}, or earlier proposals, \eg by Bortzmeyer~\cite{draft-bortzmeyer-dns-json}.
Instead of mapping the DNS fields as defined in RFC~1035~\cite{RFC-1035}, it uses arrays and their order to encode DNS messages.
Since it was the most recent version when we started, our evaluation is based on version 6 of the personal draft~\cite{draft-lenders-dns-cbor}.
Consequently, our evaluation of Packed CBOR is based on version 9 of that working group draft~\cite{draft-ietf-cbor-packed}.
DNS queries are CBOR nested arrays, \ie an array contains an array that represents the question section and up to two optional arrays that represent the authority and additional sections.
Using predefined rules, some information does not need to be carried explicitly.
For example, the number of elements in the outer array give the hint whether the authority or additional sections are present, with the authority section given priority on elision.
Within a question, record type and record class can be elided, \ie not included into the array, if they belong to the AAAA record type or IN class, respectively.

The \emph{packed} media type extension \texttt{;packed=1} utilizes the Packed CBOR format~\cite{draft-ietf-cbor-packed}.
It adds the compression of suffix names and IP address prefixes.
Redundant numbers can also be compressed to a degree, see \cref{sec:background}.
However, encoding with Packed CBOR is more complex:
When packing an existing CBOR-represented object, the encoder needs to keep state of all values, byte string prefixes, and text string suffixes within the CBOR object.
It needs to keep track how often each value appears in the object and how large the reference to the value is compared to the value itself in bytes.
This kind of state-handling is too expensive for the original constrained IoT use case of the DNS format, which is why Packed CBOR is only used for the responses from the typically much more powerful DNS server.
The pre-generating and storing of DNS messages is also not feasible on client platforms which only might have a handful of kilobytes of storage.
Since all values need to be known, Packed CBOR has another disadvantage in that it cannot be sent in-line, \ie during the generation of the object representation.
An advantage of unpacked CBOR over classic compression approaches such as GZip or Brotli.
Lastly---in a similar vain to classic compression---it is not universal; when the object contains no redundancy at all, \emph{packed} adds unnecessary overhead of an empty packing table.

In the following, we use \emph{unpacked} for the base \texttt{application/dns+cbor} format and \emph{packed} for the \texttt{application/dns+cbor;packed=1} format.

\begin{figure*}
  \centering
  \subfloat[Matching SLDs/TLDs.]{\includegraphics{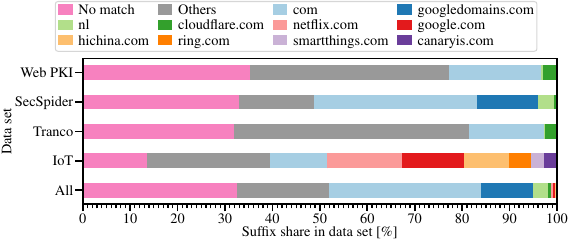}\label{fig:eval-c4d:sld-names}}%
  \subfloat[Matching SLD/TLD categories.]{\includegraphics{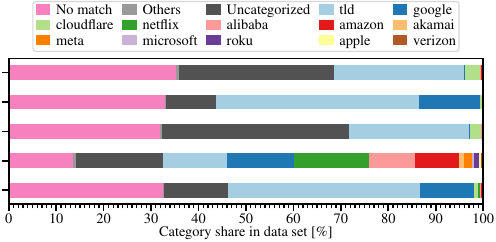}\label{fig:eval-c4d:sld-categories}}
  \caption{Distribution of SLDs and TLDs of matching name suffixes, ignoring equal names, in DNS messages and their category according to V2Fly list~\cite{v2fly}. ``Other'' summarizes all entries with a share below $2.0\%$ and $0.3\%$, respectively, in each data set.}%
  \label{fig:eval-c4d:slds}
\end{figure*}

\paragraph{Taxonomy}
We extend the taxonomy used in \cref{sec:eval-j2c} to consider both the size of the encoded CBOR object for the \textbf{tier} determination and the \textbf{content types} byte strings (\emph{binary}) and CBOR tags (\emph{taggy}).
Unless they already have other definitions, \eg boolean literals or floats, we classified objects consisting of major type 7 as either \emph{taggy}, when a \emph{simple value} is used as a reference pointer within \emph{packed} ($N<16$)~\cite{draft-ietf-cbor-packed}, or \emph{numeric}, for all other cases.

Based on our extended taxonomy, we classify data using \emph{unpacked} and \emph{packed} formats. %
$98.5\%$ of all CBOR objects in our data set are \emph{tier 1}.
$5.9\%$ of the \emph{packed} objects are in \emph{tier~2}.
Only $0.02\%$ of all objects are in \emph{tier 3}. %
Larger objects in \emph{packed} are more likely because \emph{packed} only encodes responses, which are typically larger than queries.
This also leads to diverging results for content types.
$86.3\%$ of \emph{unpacked} objects are \emph{textual}, $45.7\%$ of \emph{packed} objects \emph{binary} and $40.7\%$ are \emph{numeric}. %

Queries mostly consist of names, which are encoded as text strings.
Responses, on the other hand, contain mostly IP addresses or raw resource records---both encoded as byte strings---or numbers, \eg for TTL or to identify record types.
$10.1\%$ of \emph{packed} objects are \emph{taggy} mostly due to packing table references.
Over $99.4\%$ of \emph{unpacked} messages are \emph{non-redundant}. %
For \emph{packed}, which aims to reduce redundancy, it is even more than $99.8\%$.
$57.0\%$ of \emph{packed} is classified as nested due to the outer array to prepend the packing table. %

\paragraph{Gain and byte savings}
In \cref{fig:eval-c4d}, we show the byte savings $b$ and gain $g$, as defined in \cref{eq:gain,eq:savings} in \cref{sec:eval-j2c}. %
``Minified JSON size'' is replaced in this section by the size of the classic uncompressed DNS message in its wire-format, \eg as it would be transferred with DNS over UDP~\cite{RFC-1035}, DNS over HTTPS with the POST method~\cite{RFC-8484}, and DNS over CoAP~\cite{RFC-9953}.
We use the responses that cannot be mapped to a query to test gain and byte savings when the question section is not elided in the response, see \emph{responses (w/ question)} in \cref{fig:eval-c4d}.
For all other responses, we use the question from that query to confirm that the question is to be elided, \emph{responses (w/o question)} in \cref{fig:eval-c4d}.

Up to $202$ bytes can be saved for \emph{queries} due to traffic related to Multicast DNS (mDNS) for DNS service discovery (DNS-SD).
In mDNS, the answer section is allowed to contain known answers~\cite{RFC-6762}, but in the CBOR format the answer section in a query is always elided.
The minimum of $-6.3\%$ gain is caused by DNS-SD ANY queries, which contain SRV record names in the authority section.
The \emph{unpacked} format may actually add bytes because \emph{unpacked} has no name compression, leading to a byte saving minimum of $-1{,}181$~bytes for \emph{unpacked responses (w/o question)} in our data set.
We found a message that contains the same name suffix of $13$~characters $301\times$ in the answer section.
\emph{Packed}, however, may also cause penalties using the CBOR format.
A negative gain of $-130\%$ is caused since SOA records and their names are encoded as byte strings.
As such, suffixes of large MNAME and RNAME fields in a SOA record~\cite{RFC-1035} cannot be added to the packing table.

When there is significant compression potential, \eg redundant IP address prefixes, \emph{packed} really shines.
We found a case with $1{,}454$ A records in the answer section, leading to a gain of $95.5\%$.
This reduced the size of the original DNS message from $23{,}298$ bytes to $14{,}547$ bytes \emph{unpacked}---since names that match the question name were elided---and again to $10{,}214$ bytes when \emph{packed}.

\subsection{Potentials for Optimization by Sharing Common IP Prefixes and Name Suffixes}\label{sec:eval-c4d:comp-pot}

\noindent Our analyses of gain and byte savings in \cref{sec:eval-c4d:dns+cbor} show that the lack of name compression is a drawback of the \emph{unpacked} format when name suffixes repeat.
Elision when names match to the queried name does not help alone but \emph{Packed} mitigates these disadvantages.
However, \emph{packed} encoding is more complex, especially on constrained devices.
It cannot easily be used for in-line sending, and is not universal (see \cref{sec:eval-c4d:dns+cbor}).
To understand opportunities for improvements, we now analyze both the potentials of eliding repeating suffixes from names and prefixes from IP addresses within DNS~responses.
We exclude DNS~queries from this analysis as they only contribute very limited amount of name and IP~address~material.

To determine a common name suffix, we apply two approaches: \emph{byte-wise} and \emph{component-wise} matches.
A \emph{byte-wise} common suffix consists of those characters that overlap most when comparing two names starting from the right side (\eg \texttt{service.example.com} and \texttt{we-sell-}\linebreak\texttt{ice.example.com} share the suffix \texttt{ice.example.com}).
A \emph{component-wise} comparison identifies the name components that match (\eg in our example the components \texttt{.example.com}).
When we determine common name suffixes and address prefixes, we compare pairwise all names and addresses included in a DNS~response.

\begin{figure*}
  \centering
  \includegraphics{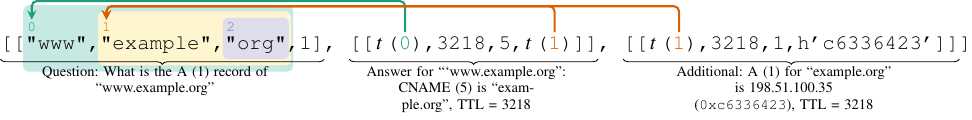}%
  \caption{Example for \emph{Component Referencing} in CBOR diagnostic notation. $t$~is the reference tag.}%
  \label{fig:comp-ref-example}
\end{figure*}
\begin{figure*}
  \centering
  \adjustbox{clip=True, trim=0 10.6em 0 0}{\includegraphics{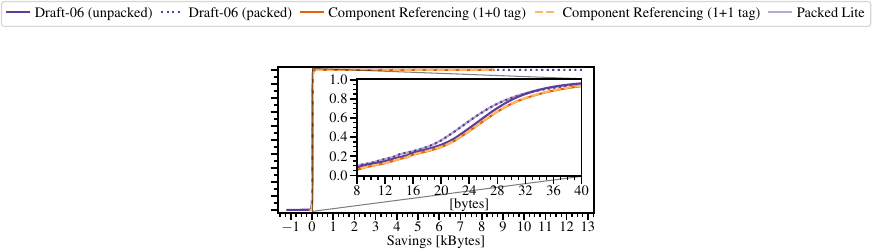}}\\[-1em]
  \subfloat[Queries.]{\includegraphics{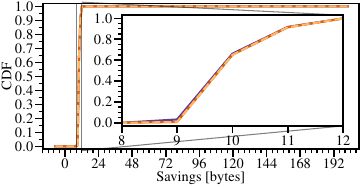}\label{fig:eval-c4d:name-comp-savings:queries}}\hspace{1mm}%
  \subfloat[Responses w/o question.]{\includegraphics{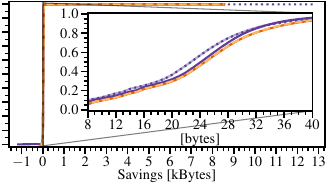}\label{fig:eval-c4d:name-comp-savings:resp_woq}}\hspace{1mm}%
  \subfloat[Responses w/ question.]{\includegraphics{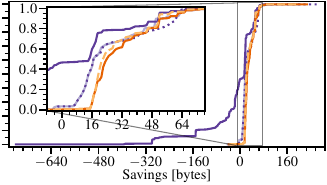}\label{fig:eval-c4d:name-comp-savings:resp_wq}}\hspace{1mm}%
  \caption{Byte savings for our name compression proposals over the classic DNS message format (\texttt{application/dns-message}). ``Draft-06'' denotes the original base format analyzed in \cref{sec:eval-c4d:dns+cbor}. Note that every plot has its own x-axis scaling for better visibility.}%
  \label{fig:eval-c4d:name-comp-savings}
\end{figure*}

\paragraph{Types of name suffixes}
To better understand how likely specific name suffixes match, we determine the second-level domain by identifying the first name label left to the top-level domain based on the Mozilla Public Suffix List~\cite{mozilla-psl}.
If two suffixes in our data set do not match, we classify the domain as \emph{no match}, which also includes name pairs that have no common suffix.
We categorize the resulting SLDs and TLDs using the V2Fly domain list~\cite{v2fly}, a community-maintained domain list that groups them by companies, business type, topics, and other points of interest.

\cref{fig:eval-c4d:slds} exhibits the share of matching SLDs/TLDs and categories.
$32.5\%$ of all name pairs did not match.
We were unable to categorize $13.6\%$ of matches, mostly because they did not belong to a significant domain.
The \texttt{.com} TLD matches most with $31.8\%$. %
In the \emph{IoT} data set, most of the names are registered under common domains, only $0.7\%$ contribute to the ``other'' category. %
\emph{Tranco}, \emph{SecSpider}, and \emph{WebPKI} have a stronger preference towards communications with Google or Cloudflare, though this overall share only comprises $12.2\%$ in total.

\paragraph{Name suffix length}
We also look into common suffix lengths, both \emph{name-label-wise} and \emph{byte-wise} (not shown).
Most query types share 4 bytes~(median). %
Responses for A record queries contribute $2.7\%$ to all name pairs.
Additional to the A record, those responses often contain records that tend to carry names (\eg SOA or CNAME), leading to a common suffix length of 14 bytes (median).
SOA records frequently cause a 2 byte \emph{byte-wise} match, \eg when they specify a ccTLDs zone.
In at least 50\% of messages, however, each suffix appears at most once (not shown).

\paragraph{IP address prefix length}
Similarly, we look into common IP prefix lengths (not shown).
It is worth noting that A (or AAAA) records might be part of an AAAA (or A) query response, \eg addresses for NS records.
Overall, the median for IPv4 prefix shares is 1 byte and for IPv6 addresses 5 bytes.
This complies with the intuition that IPv4 addresses allow for less compression potential because they are shorter.
In the case of IPv6 addresses, there is significantly more potential, since $\approx 30\%$ $\left(\frac{5\ \text{bytes}}{16\ \text{bytes}}\right)$ per IPv6 address can be elided.

\subsection{Name Compression in Unpacked application/dns+cbor}\label{sec:eval-c4d:name-comp}
\noindent We found that \emph{unpacked} can gain an advantage when compressing names or IP addresses, see \cref{sec:eval-c4d:comp-pot}.
To leverage these potentials, we propose two possible schemes for name compression in the \emph{unpacked} format as names offered the largest potential for compression.
\one Name \emph{Component Referencing} and \two \emph{Packed Lite}.
Full \emph{packed} would still be applicable for value and address prefix compression at a server.

\paragraph{Optimization 1: Component Referencing}
Inspired by classic DNS~\cite[\S 4.1.4]{RFC-1035} and in contrast to the evaluated version 6 of our draft~\cite{draft-lenders-dns-cbor}, names in this scheme are not encoded as text strings but split into components and then considered as a sequence of text strings. %
During construction, these text strings are indexed with an integer depth-first.
Each CBOR sequence of name components, including each of its suffixes, is kept track by the index of the first string.
Whenever one of the suffixes needs to be encoded, the encoder inserts the index instead, marked by a special reference tag. %
The decoder then can just count the text strings while decoding.
If the decoder encounters a reference tag, it appends the text strings to the name by jumping to that referenced text string, appending any following text string, and dereferencing any further reference tags after those.
The decoder stops appending text strings when there are no text strings or reference tags anymore.
An example encoding of a DNS message with \emph{Component Referencing} can be seen in \cref{fig:comp-ref-example}.
As long as name components are shorter than 24~bytes, this name encoding needs at most the same space as encoding the full name as a text string.
In our analysis, we used two types of reference tags: A \emph{1+0 tag}, which encodes to 1 byte in total, and a \emph{1+1 tag}, which encodes to 2 bytes (see~\cref{sec:background}).
\begin{figure*}
  \centering
  \subfloat[Encoder runtime.]{\includegraphics{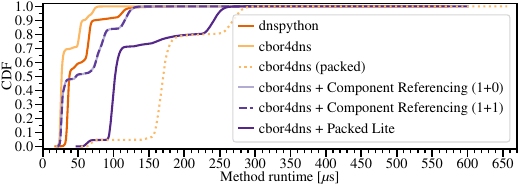}\label{fig:impl:encode}}\hspace{1mm}%
  \subfloat[Decoder runtime.]{\includegraphics{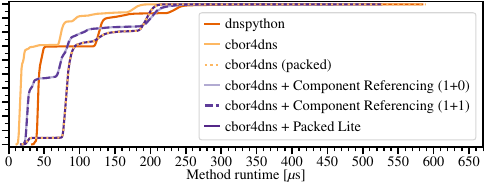}\label{fig:impl:decode}}\hspace{1mm}%
  \caption{Encoding and decoding runtimes for DNS messages that generate small, binary or textual, non-redundant, and flat base \texttt{application/dns+cbor} messages for \texttt{dnspython} and (extended) \texttt{cbor4dns}.}%
  \label{fig:impl:runtime}
\end{figure*}

\paragraph{Optimization 2: Packed Lite} 
To simplify \emph{packed} complexity, we only add suffixes to text strings to the packing table.
This lean extension allows the encoder to only keep state of the text string suffixes, which is of similar complexity compared to \emph{Component Referencing}.
However, similar to the full \emph{packed} format, the object cannot be encoded in-line.
This compression scheme would ease implementation, though, as only one compression needs to be implemented in a full encoder, \emph{Packed Lite} would only restrict which type of value is added to the packing table.
Hence, it serves as a performance baseline for the existing name compression in \emph{packed}.

\paragraph{Gain and byte savings}
In \cref{fig:eval-c4d:name-comp-savings}, we compare the byte savings of the base \emph{packed} and \emph{unpacked} format (\emph{Draft-06}) for \emph{queries} (\cref{fig:eval-c4d:name-comp-savings:queries}), \emph{responses w/o question} (\cref{fig:eval-c4d:name-comp-savings:resp_woq}), and \emph{responses w/ question} (\cref{fig:eval-c4d:name-comp-savings:resp_wq}), using the grouping as in \cref{sec:eval-c4d:dns+cbor}.
For brevity, we do not show the plots for gain.
For the queries, we only look at \emph{Component Referencing} as our \emph{Packed Lite} implementation only modified the full \emph{packed} implementation within \texttt{cbor4dns}.
Here, \emph{Component Referencing} shows little difference compared to the \emph{Draft-06}, as most of them only contain a single name in the question section.
For responses, \emph{Packed Lite} closely follows the original \emph{packed} format.
\emph{Component Referencing (0+1)} performs very similar in \emph{Responses (w/ question)}.
\emph{Draft-06} only saves $110$ bytes at maximum ($47.8\%$ gain) and \emph{Packed Lite} $203$ bytes ($49.2\%$ gain), while \emph{Component Referencing} saves $226$ bytes ($52.6\%$ gain), \ie $116$ bytes more than \emph{Draft-06}.

\section{Evaluating the Name Compressors}\label{sec:impl}

\noindent To evaluate their performance and complexity we provide an implementation for the name compression using \emph{Component Referencing} and \emph{Packed Lite}.
Our implementations extend our \texttt{cbor4dns} tool~\cite{cbor4dns}, which was already part of our analysis in \cref{sec:eval-c4d}.
For \emph{Component Referencing}, we add a Python map as state, which we update with the name components and their indices during encoding and decoding.
For \emph{Packed Lite}, we add a flag to the encoder to skip adding anything but text strings or references to text strings to the packing table.
The decoder remains unchanged for \emph{Packed Lite}.

An implementation of name compression in the CBOR-based DNS format should be as fast as a classic DNS message composer and parser.
It should also be suitable for the original constrained IoT use case, \ie fit into a small ROM and memory footprint.
We compare runtimes of the new encoder and decoder with the original base implementation of \texttt{cbor4dns} and the \texttt{dnspython} toolkit for classic DNS messages.
While the new feature  adds overhead to the base implementation, we confirm that both proposals are less complex than the full \emph{packed} format.
We also examine the suitability for the IoT use case of our \emph{Component Referencing} extension by evaluating the build sizes of an implementation for the RIOT operating~system~\cite{bghkl-rosos-18}.

For our runtime analysis, we compare each extension to the \texttt{dnspython} DNS message encoder and decoder v2.4.2~\cite{dnspython}, and the original version of \texttt{cbor4dns}.
We query our IoT data set (see \cref{sec:eval-c4d}) as this provides the largest name diversity.
For consistent results, we restrict the data set to DNS messages that generate non-nested, non-redundant, and small \emph{unpacked} CBOR objects that largely contain text or byte strings.
This selection comprises $1{,}040{,}386$ queries and $456{,}819$ responses.
For each DNS message we use the format of \texttt{dnspython} as encoder input, which \texttt{cbor4dns} also uses, and the respective binary form as decoder input.
We measure the runtime for 100 runs of each function on our AMD EPYC 7702 processor.
The results for the average over those 100 runs can be seen in \cref{fig:impl:runtime}.

\emph{Component Referencing} performance falls between the packed formats and the original base formats of \texttt{dnspython} and \texttt{cbor4dns}, with median encoder time of 47~$\mu$s and decoder time of 69~$\mu$s.
There is no significant difference between the \emph{1+0 tag} and \emph{1+1 tag} variant.
The feature-reduced \emph{Packed Lite} encoder performs about 63~$\mu$s better than the full \emph{packed} encoder in the median, while for the unmodified decoders both gather at median 84~$\mu$s.
For \emph{Component Referencing} on the RIOT operating system for the constrained IoT, the response decoder fits into 314 bytes of ROM, which makes it very suitable for the constrained IoT.
Name compression for the encoder is not needed, as in the RIOT use case the query just consists of the question section with one question.

\section{End-to-end Validation}\label{sec:eval-e2e}

\begin{figure}
  \centering
  \includegraphics{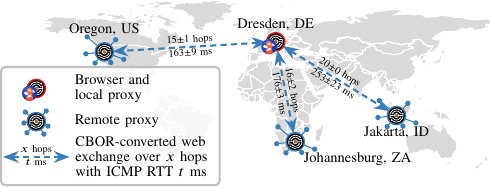}
  \caption{%
  Setup for end-to-end validation. The proxies convert uncompressed JSON and DNS messages to CBOR but leave everything else intact.
  The browser audits the websites of the top~100 and bottom~100 domains from the Tranco Top 1M list~\cite{lvtkj-tranco-2019} from April 13, 2026, using Lighthouse~\cite{google-lighthouse}.
  }\label{fig:e2e-setup}
\end{figure}

In this section, we evaluate, under real-world conditions, the end-to-end performance of browser-to-server communication when using the CBOR proposals introduced in \cref{sec:eval-j2c,sec:eval-c4d}.
We focus on the performance perceived by mobile Web users requesting distributed content, since they benefit most from a leaner and faster Web experience.

Our analysis is challenged for two reasons.
First, neither Web browser nor server deployments currently provide advanced CBOR support.
Second, the caching infrastructure of Content Delivery Networks~(CDNs) may relocate content.
Therefore, we use local and remote proxies that transparently convert Web traffic between JSON or DoH and CBOR for browsers and distributed Web servers.

\subsection{Setup}
\cref{fig:e2e-setup} visualizes our setup.
We deploy the client in our university network and remote proxy instances in three different regions using Google Cloud Platform:
Oregon, USA (US, \texttt{us-west1}),
Johannesburg, South Africa (ZA, \texttt{africa-south1}), and
Jakarta, Indonesia (ID, \texttt{asia-southeast2}).
Edges show the average distances in IP hops and the ICMP RTT for each location.

We built our CBOR proxy based on \emph{mitmproxy}~\cite{mitmproxy}.
This implementation decodes \emph{every} request and response as JSON unless the payload is a DNS~message (\ie the \texttt{Content-Type} field is set to \texttt{application/dns-message}).
Only parsable objects (with at least one key-value pair) and arrays (with at least one element) are converted to CBOR\@.
DNS~messages are converted with \emph{Component Referencing (1+0 tag)} and are left~unpacked.

To evaluate perceived performance, we use Google Lighthouse, a popular website performance auditing tool integrated into Google Chrome v147~\cite{google-lighthouse}.
Google Lighthouse provides performance measures but must be run in graphical mode, which is why we run the browser on our site.
To approximate the Internet connectivity typical for households with low Internet connectivity~\cite{dmczo-ccsr-21,mhrr-rttdi-24}, we throttle the download speed to 50 MBits/s and the upload speed to 20 MBits/s based on the Lighthouse mobile profile.

To account for side effects due to caching, we first pre-populate remote CDN caches by sending unmodified requests from the browser through the local proxy to the remote proxies.
This run is discarded from further analysis.

In our measurements, we consider the top~100 and bottom~100 domains from the Tranco Top~1M list~\cite{lvtkj-tranco-2019}.
To enable a comparative analysis, we evaluate the same domain names both with and without CBOR encoding.
To enhance statistical robustness, we repeat each run 30~times, grouping the repetitions into daily batches of five and alternating the remote proxy location for each batch.
This approach allows us to capture results across different times of the day and weeks for all locations.

\subsection{End-to-end Measurements Results}
We use the Tranco list from April 13, 2026, and run our measurements between May 3 and May 23, 2026.
Overall, we query 200 websites from three locations with and without CBOR encoding, repeating each query 30 times, which yields a total of 36{,}000 website audits.

Only 62.0\% of all queries result in a successful audit (mean 124.9 websites per run for US, 121.1 for ZA, and 123.3 for ID).
17.9\% requests lead to an interstitial page in Chrome (usually due to a name resolution error), 12.7\% resulted in HTTP errors (\eg not found, forbidden), 6.9\% resulted in a \texttt{NO\_FCP} error (usually due to a response timeout), and 0.5\% other failures (\eg not rendering to HTML or hanging JavaScript).
Some problems relate to the Tranco list, which only contains second-level domains (SLD).
An error can occur when the original domain name (\eg \texttt{www.example.com}) resolves but its SLD (\eg \texttt{example.com}) does not.
Moreover, not every website provides access to the root path (\eg \texttt{example.com/} does not resolve but \texttt{example.com/index.html} does).
Nevertheless, domain names from the Tranco~list are common in Web measurements.

Our proxies gathered $\approx 57{,}500$ HTTP requests and $\approx 56{,}000$ HTTP responses.
Of those, about 23.7\% are compressed JSON requests---using \eg GZip or Brotli---and 3.7\% compressed JSON responses, which we do not convert to CBOR\@.
When converting uncompressed data to CBOR, 3.8\% were requests and 2.7\% were responses.
83.6\% of the total requests were converted DoH queries and 92.2\% were converted DoH responses.
On average, we reduce the payload by 1,026.7~bytes per website and audit for JSON messages and 57.0~bytes for DNS messages.
We tried to equalize the handler times for both unconverted and converted runs.
Despite this, converted runs take on average 22.4~ms per website audit longer than the unconverted runs.

\begin{figure}
  \centering
  \includegraphics{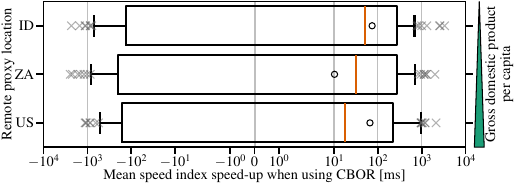}
  \caption{%
    Speed-up of the mean Lighthouse speed index, \ie the difference between runs without and with CBOR, per website for end-to-end validation over 30 runs. %
    The boxes extend from the 25th to the 75th percentile.
    The median is marked with the orange line within the boxes.
    The circles mark the arithmetic mean of each data set.
    The whiskers extend from the 5th to the 95th percentile, outliers beyond those are marked with {\color{Dark2-H}$\times$}.%
  }\label{fig:e2e-results}
\end{figure}

\cref{fig:e2e-results} shows our results for the mean difference in speed index over all runs between unconverted and converted runs (\ie the speed-up) measured by Lighthouse.
The speed index is the area over the curve of the percentage of completeness of a loading webpage over time, \ie a lesser speed index means a faster completed site.
Other metrics, such as First or Largest Contentful Paint, are unsuitable for our purposes because they primarily measure the rendering of visual elements (\eg images or videos).

Despite our efforts to reduce noise through warm-up phase and multiple runs, our results include large standard deviations.
This is mostly due to \one pages loading different content parts first at every reload and \two highly fluctuating RTTs of upstream networks.
To reduce outliers, we exclude all data points with an absolute $z$-score larger than 5.
We choose this high $z$-score to only filter the most extreme cases.

The mean for US is 66.0~ms, for ZA 10.3~ms, and for ID 73.9~ms.
The median for US is 18.0~ms, for ZA 31.7~ms, and for ID 50.4~ms.
Both median and mean are positive and comply with our results presented in \cref{sec:eval-j2c:times}.
This shows an overall improvement in page load times when using CBOR in realistic scenarios, also indicating that countries with less well-equipped Internet infrastructure benefit more.

\section{Discussion}\label{sec:discussion}

\paragraphS{Why are even small byte savings of web content important}
Milliseconds matter.
Web content of today is not only consumed over broadband access networks but also over narrowband or intermittent connectivity.
Rural areas of the Western world~\cite{tsfw-ebass-13}, developing countries~\cite{zcpas-dwlg-14,caazr-twwwd-23}, and entire continents~\cite{arctic-blog} live with a slow Internet, see also \cref{sec:eval-e2e}.
Anyone who ever tried to access the Web in these regions knows these issues.
In consequence, users do not consume large media files but require patience when downloading plain text.
Any approach to reduce loading times---even smaller ones in the magnitude of a few handful of bytes as presented in \cref{sec:eval-j2c} and \cref{sec:eval-e2e}---will increase accessibility and improve usability of the Web for these communities.
Furthermore, mobile travelers around the world repeatedly experience low, intermittent connectivity and need to load web pages often within short intervals of Internet access.
These users will experience faster downloads from the border between success and failure.
Hence, we believe that faster web access, including leaner DNS resolution and smaller web content, are beneficial for the majority of users, regardless of existing network infrastructure.

\paragraphS{Should CBOR be used instead of JSON}
We have identified benefits when converting to CBOR in \cref{sec:eval-j2c} and \cref{sec:eval-e2e}. This alone is an argument for CBOR over JSON\@.
Both encoding and decoding take roughly the same time in our disadvantaged evaluation setup.
A further argument is the extended feature set such as byte strings and tags.
We also found, however, cases where CBOR is larger, especially when encoding floating point numbers, which can be very short in JSON\@.
Using alternative representations, such as strings or just reducing precision in CBOR could further outperform JSON\@.
Native browser support is currently limited, but there are JavaScript libraries available, as well as WebAssembly implementations in other programming languages such as Rust~\cite{cbor-js,cbor-wasm,purewasm-cbor}.
Compared to plain JSON, CBOR is not human-readable, but JSON is often sent in a minified format.
We argue that minified JSON is at least as hard to parse for the human eye as, for example, \texttt{\textbackslash{xXX}}-encoded CBOR\@.
To improve readability in CBOR for humans, the CBOR diagnostic notation could be a fallback.
An extended form of that notation is currently in the process of standardization in the IETF~\cite{draft-ietf-cbor-edn-literals}.
As various encoder/decoder implementations already exist for all these formats in different programming languages, we deem the technical challenge of a switch-over low.
The largest task would be to integrate CBOR encoders/decoders in common web front- and backend frameworks and CBOR could be used instead of JSON for, \eg classic AJAX or REST tasks.

\paragraphS{Should there be a new DNS message format}
The CBOR format for DNS messages comes timely.
There are loose discussions within the IETF about a general new DNS wire format~\cite{ietf-118-hackathon}.
We showed that the \emph{packed} version has large advantages over the classic DNS format, with byte savings of up to 12.8~kBytes in DNS responses.
The \emph{unpacked} version provides less potential in terms of compressing names but helps to reduce other elements in a DNS~response.

Changing the DNS message format is challenging because vendors and operators are afraid of breaking existing services.
However, especially over DoH, changes in format already have proven possible, \eg for the public resolver by Google~\cite{google-dns-json-api}.
Such deployments, however, often use very verbose JSON formats, \ie they can benefit even more from converting to \texttt{application/dns+cbor} when compared to the classic DNS format.
Our runtime analysis in \cref{sec:impl} shows that the CBOR encoder/decoder is comparable to a similar encoder/decoder for the classic DNS format.
It is worth stressing again that CBOR implementations exist in various programming languages, which helps to avoid implementation mistakes from the past~\cite{RFC-9267}.
As such, in terms of technical challenges, once more implementations of the CBOR format exist, we deem the change-over to be a simple plug-and-play, once browser vendors and DNS resolvers provide implementations supporting CBOR on DNS.

\paragraph{Deployability}
Replacing JSON by CBOR on the front- and backend side only requires changes to Web libraries.
To support CBOR-based DNS, browser vendors and DNS resolvers need to implement changes.
Alternatively, our proxy setup demonstrates an option for incremental deployment, see~\cref{sec:eval-e2e}.

The two proposed CBOR formats, generic object and DNS message representation, were specifically designed for low-resource environments.
Consequently, users of low-end mobiles could install a proxy locally.
The added complexity of a proxy is justified when the link quality is very bad and a more optimized implementation than our prototype is used.
Our evaluation shows that it still yields a net benefit.

\paragraph{End-to-end evaluation focuses on a client deployed in Germany}
Lighthouse requires full rendering of content in Chrome.
As a result, we were unable to deploy the client-side component in other regions than Germany---long-distance, remote screen mirroring would disturb the measurements.
Therefore, we requested content from our location through remote servers in the US, South Africa, and Indonesia.
We argue, however, that our results closely approximate the performance clients would experience when deployed in any of those regions and requesting content from Germany, because the RTT between remote and local setup can be assumed to be roughly symmetric.

\paragraph{Impact in the IETF}
Three results of this paper have already been adopted and integrated into the Internet draft for \texttt{application/dns+cbor}~\cite{draft-lenders-dns-cbor}.
The principle idea of \emph{Component Referencing} is adopted for three reasons.
\one~It is easier to implement on constrained hardware,
\two~it provides in-line encoding support, and
\three~it performs at least comparable, if not better, than \emph{Packed Lite}.
In the final version, however, we use a dedicated table setup tag that generates an implicit Packed CBOR packing table to avoid IANA~allocations of a reference tag.
The packing table allows for the usage of simple values instead of tags for the component referencing, saving even more bytes compared to \emph{Component Referencing (1+0)} as it only requires 1 byte per reference instead of~2.
Implementation-wise it is equivalent to \mbox{\emph{Component~Referencing}}.

We found that binary encoding of structured resource records that contain names such as SRV or SOA does not allow for name compression even with \emph{packed}.
Therefore, we introduced an array-based structure for the relevant records specified in RFC~1035~\cite{RFC-1035} to our draft~\cite{draft-lenders-dns-cbor}.
According to our evaluation in \cref{sec:eval-c4d}, a median of $10$ bytes is potentially saved per record with name compression.

Finally, the current version of our draft proposes to include the answer section in queries at the highest elision priority because we showed that queries currently drop the answer section.
This part of the specification provides support for DNS-SD and Multicast DNS (mDNS), helping name-based service discovery in IoT scenarios.

\paragraphS{How can CBOR contribute even more to the Web}
Replacing the serialization format of JSON objects is simple and brings benefits.
A CBOR-based format for DNS messages exists already but needed improvement.
This is why we restricted our evaluation to those two dynamic content formats.
However, recent research such as MAML~\cite{pavsz-tfwdr-24} reconsiders web specifications to simplify webpages based on a JSON-based serialization.
New web languages such as these will also benefit from CBOR\@.

\paragraph{CBOR for compression in LLM scenarios}
While this paper was being written, large language models (LLMs) attracted considerable attention.
Recent work~\cite{lxsdp-qcile-26} suggests that constrained applications exchange models and features with a more powerful edge device to enable LLMs in constrained scenarios.
To account for low resources, model compression is proposed.
Since LLMs and deep learning algorithms are mostly vectors and matrices---\ie arrays and arrays of arrays---of floating point numbers (floats), this is, among other techniques, achieved by minimizing the bit-width of floats~\cite{lxxsr-lmela-25,lxsdp-qcile-26}.
When not accounting for decimal fractions (tag~4) or bigfloats (tag~5), CBOR only knows three widths for floats: 16 bits (half precision), 32~bits (single precision), and 64 bits (double precision)---all on a byte-aligned level~\cite{RFC-8949}.
Model compression, on the other hand, is much more varied, compressing floats down to even a single bit.
As such, representing compressed models in CBOR has little benefit.
However, compared to encoding compressed models in \texttt{base64} and then carrying them in, \eg JSON, CBOR provides more lightweight byte strings, see \cref{sec:eval-j2c:gain-savings}.

\section{Conclusion and Future Work}\label{sec:conclusion}
\noindent In this paper, we showed that the World Wide Web can benefit from CBOR\@.
Encoding objects in CBOR instead of JSON can save more than 15.2 kBytes (18.8\%) on average per object.
A more concise CBOR format of DNS, which is a notable contributor to the initial loading delay, is already under discussion in the IETF\@.
We contributed two name suffix compression schemes for the CBOR format, \emph{Component Referencing} and \emph{Packed Lite}, following our analysis for compression potentials.
\emph{Component Referencing} showed slightly better performance and a leaner implementation.
Our implementation of a query encoder and a response decoder with \emph{Component Referencing} for embedded systems fits within 2.1 kBytes.
Thereof, the \emph{Component Referencing} parser only required 314 bytes.
Beyond name compression, we derived further recommendations for improvement to the base format.
We adopted most of these improvements in our ongoing IETF work~\cite{draft-lenders-dns-cbor}.
We demonstrated that replacing JSON and classic DNS by CBOR reduces the time to completion of a web page by median 18.0 to 50.4~ms depending on RTT.

This work opens three future research directions.
First, the evaluation of the {application/dns+cbor} format should be extended to DNS traffic from the Internet backbone and used to optimize messages even further.
Second, the potentials of converting further common web formats, \eg HTML via MAML, to CBOR should be investigated.
Third, protocol techniques and methods designed for the constrained IoT should
be analyzed with the objective of optimizing the global Web to reduce digital inequality.

\paragraph{Artifacts}
The software artifacts of this work, \ie implementation and experiment evaluations are available under \textbf{\href{https://doi.org/10.5281/zenodo.21790597}{doi:10.5281/zenodo.21790597}}.
The data artifacts, \ie the data sets collected and derivates of these data sets are available under \textbf{\href{https://doi.org/10.25532/OPARA-1530}{doi:10.25532/OPARA-1530}}.

\section*{Acknowledgments}
\noindent We would like to thank Gareth Tyson, Christian Amsüss, Lucas Vogel, and the anonymous reviewers for their suggestions and comments on this article.

\label{lastpage} %

\bibliographystyle{IEEEtran}
\bibliography{./local,rfcs,hypermedia,ids,internet,iot,layer2,own,security}

\balance

\appendix

\subsection{CBOR Encoding and Major Types}\label{sec:cbor-semantics}

\begin{figure}[!htbp]
  \begin{center}
    \includegraphics{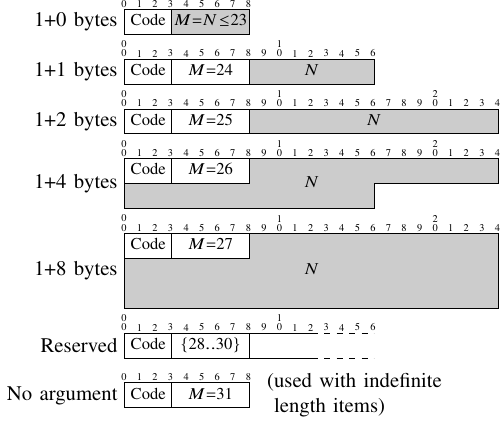}
    \caption{Headers of CBOR elements. The type code and the indicator $M$ determine the argument $N$.}%
    \label{fig:cbor-type-head}
  \end{center}
\end{figure}

\noindent For each CBOR element the major type and the 5-bit indicator $M$ determine the meaning of $N$, see \cref{fig:cbor-type-head}. There are 8 major types, identified by a 3-bit code:

\begin{enumerate}
\setcounter{enumi}{-1}
\item \emph{Unsigned integers} encode integers in the range from $0$ to $2^{64}-1$ inclusive. $N$ is the value of the integer.
\item \emph{Negative integers} encode integers in the range from $-2^{64}$ to $-1$ inclusive. $-1-N$ is the value of the integer.
\item \emph{Byte strings} encode strings of binary data. $N$ is the number of the raw bytes which follow after the CBOR type header.
\item \emph{Text strings} encode UTF-8 strings of binary data. $N$ is the length of the string in bytes which follows after the CBOR type header.
\item \emph{Arrays} are sequences of other elements. $N$ is the number of elements in the array that follow after the CBOR type header.
\item \emph{Maps} are sequences of key-value pairs which in turn consist of other elements. $N$ is the number of key-value pairs in the map.
\item \emph{Tags} mark other elements with special semantics. $N$ is the tag number which determines the semantic.
\item \emph{Simple values and floats} depend in their meaning heavily on $M$. For $M \leq 19$ and $M=24$, $N$ offers another integer value space in the range of $0$ to $255$ inclusive, the \emph{simple values}.
\end{enumerate}

\begin{table}[!htbp]
  \begin{center}
    \caption{Major types of CBOR and their respective meaning of $N$.}%
    \label{tab:cbor-major-types}
    \setlength{\tabcolsep}{1pt}
    \footnotesize
    \begin{tabular}{rcrp{.48\linewidth}}
      \toprule
      Major Type            & Code  & $M$               & Meaning of $N$\\
      \midrule
      Unsigned integer      & 0     & $\leq 27$         & $N$ is value of the integer.\\
      Negative integer      & 1     & $\leq 27$         & $-1-N$ is value of the integer.\\
      Byte string           & 2     & $\leq 27$         & $N$ is length of string following in bytes.\\
                            &       & $31$              & Indefinite length string.\\
      Text (UTF-8) string   & 3     & $\leq 27$         & $N$ is length of string following in bytes.\\
                            &       & $31$              & Indefinite length string.\\
      Array                 & 4     & $\leq 27$         & $N$ is the number of elements following.\\
                            &       & $31$              & Indefinite length array.\\
      Map                   & 5     & $\leq 27$         & $N$ is the number of key-value pairs following.\\
                            &       & $31$              & Indefinite length map.\\
      Tag                   & 6     & $\leq 27$         & Value of the tag = $N$.\\
      Simple value / float  & 7     & $\leq 19$         & $N$ is value of the simple values $\in \{0..19\}$.\\
                            &       & $\in \{20..23\}$  & Literals ($20$: \texttt{false}, $21$: \texttt{true}, $22$: \texttt{null}, $23$: \texttt{undefined}).\\
                            &       & $24$              & $N$ is value of the simple value $\in \{32..255\}$.\\
                            &       & $\in \{25..27\}$  & $N$ encodes IEEE 754 float ($25$: 16-bits, $26$: 32-bits, $27$: 64-bits).\\
                            &       & $31$              & Indefinite sequence ``break''.\\
      \bottomrule
    \end{tabular}
  \end{center}
  \vskip-.5em
\end{table}

Special meaning is given to some of these numbers, \eg $0 \leq M = N \leq 15$ are used as value reference to the packing table in Packed CBOR~\cite{draft-ietf-cbor-packed}, $M=20$ encodes a boolean \texttt{false}, $M=21$ a boolean \texttt{true}, $M=22$ the value \texttt{null}, and $M=23$ the value \text{undefined}.
If $M$ equals $25$, $26$, or $27$ the following 2, 4, or 8 bytes respectively encode half-precision, single-precision, or double-precision IEEE 754 floats.

$M=31$ is special in most types: for byte strings, text strings, arrays and maps it marks the following bytes of indefinite length.
These are then terminated with $M=31$ of major type~7.

\subsection{DNS Message Format}\label{sec:dns-msg}
\begin{figure}
  \setlength{\abovecaptionskip}{5pt plus 3pt minus 2pt}
  \centering
  \input{figs/dns-field-tikz}
  \subfloat[DNS query.]{%
    \centering
    \includegraphics{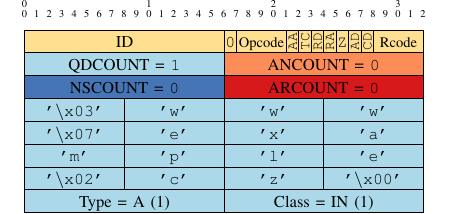}\label{fig:dns-message-format:query}%
  }\hspace{1em}%
  \subfloat[DNS response.]{%
    \centering
    \includegraphics{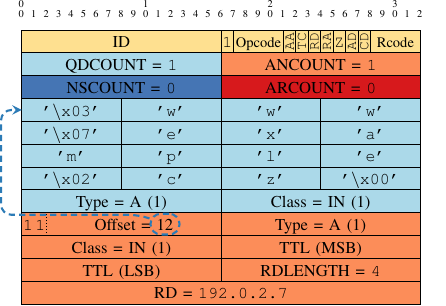}\label{fig:dns-message-format:response}%
  }
  \caption{Example DNS message format for an A record query and response for domain name \texttt{www.example.cz}.}%
  \label{fig:dns-message-format}
\end{figure}

\noindent DNS messages in the classic format specified in RFC~1035~\cite{RFC-1035} and follow-ups consist of a header of length 24 bytes and 4 sections---question (QD), answer (AN), authority (NS), additional records (AR), each of variable length.
\cref{fig:dns-message-format} depicts an example query and an example response.
The DNS header consists of 6 16-bits wide fields.
The \emph{ID} field, the \emph{flags}, the count fields for each section.
The most significant bit in \emph{flags}, \emph{QR}, determines if a message is a query (0) or a response (1).
Each section contains a number of records provided by the respective count field.
Records in the question section contain of a query name (variable length, see below), a query type (16 bits), and a query class (16 bits), and there is usually only one record in this section~\cite{RFC-9619}.
All other sections contain 0 or more resource records (RRs), which consist of an RR name (variable length, see below), an RR type (16 bits), an RR class (16 bits), a TTL (32 bits), the RR data length (16 bits, RDLENGTH in \cref{fig:dns-message-format:response}), and the RR data (length determined by RDLENGTH).
Names are encoded by components-wise, from the lowest level to the top level.
For each component, there is first a byte that determines its length with the component itself following.
After the top level component, a \texttt{\textbackslash{x00}} byte ends the name.
If the first byte starts with a \texttt{0b11000000} mask, the name is not written out.
Rather, this byte and the following form a pointer to a name component within the message.
The value of the pointer is the byte offset of the name component within the message.

\nobalance%

\section*{Author Biographies}

\begin{IEEEbiography}[{\includegraphics[width=1in,height=1.25in,clip,keepaspectratio]{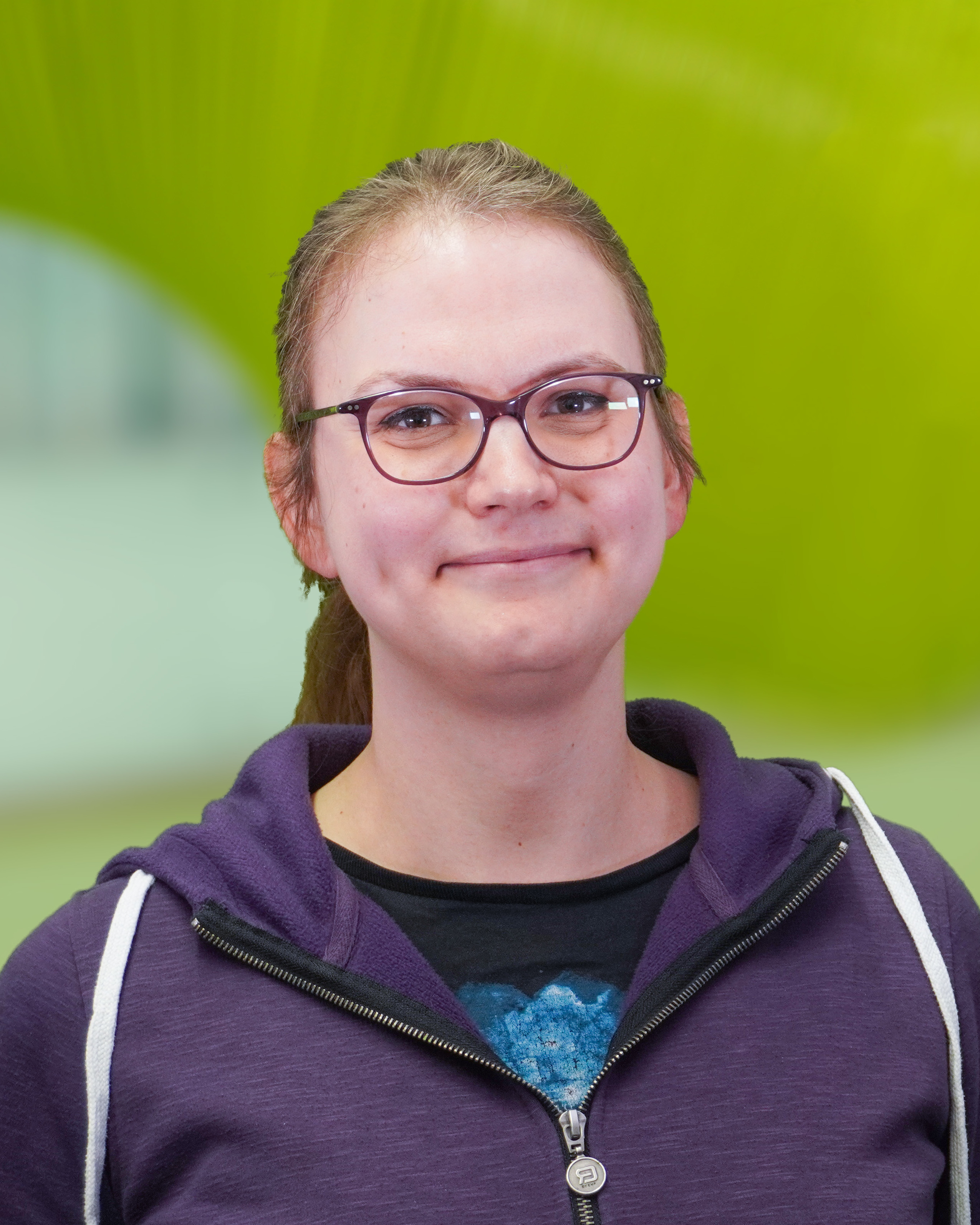}}]{Martine S. Lenders}
(Member, IEEE) is a PhD student at Freie Universit\"at Berlin (FU Berlin) and a research associate at the Chair of Distributed and Networked Systems at TUD Dresden University of Technology (TU~Dresden). Before joining TU Dresden, she graduated with a BSc and MSc in Computer Science from FU Berlin. Her research focuses on networking, privacy, and programming for the Internet of Things (IoT). Through her engagement into all things IoT and network protocols, she is also involved with the development of the IoT operating system RIOT since 2011 and its maintenance since 2013. Additionally, she is active in the Internet Engineering Task Force (IETF) since 2015 where she co-authored RFCs 9952 and 9953.
\end{IEEEbiography}

\begin{IEEEbiography}[{\includegraphics[width=1in,height=1.25in,clip,keepaspectratio]{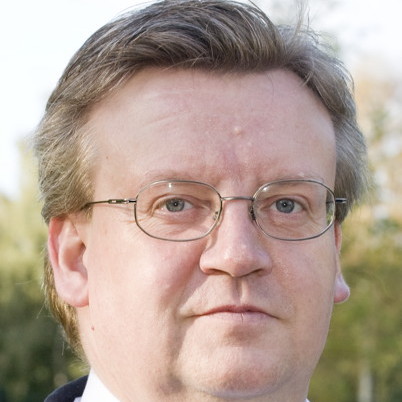}}]{Carsten Bormann}
 serves as a Honorarprofessor for Internet technology at the
Universität Bremen and is a member of its Center for Computing and
Communications Technology (TZI).
His research interests are in protocol design and system architectures for
networking. In the IETF, he mainly has been working on bringing Internet
Technology to new links, applications, or radios.
Carsten is the initiator of the Concise Binary Object Representation (CBOR)
standard, a data representation format that is quickly becoming the JSON of
resource-constrained and security-critical systems, as used for instance in the
European vaccination certificates and the ISO Mobile Driving License.
As part of his track record of Internet standardization, Carsten has
authored and co-authored 65 Internet RFCs, with more approved and
waiting for final editing.
\end{IEEEbiography}

\begin{IEEEbiography}[{\includegraphics[width=1in,height=1.25in,clip,keepaspectratio]{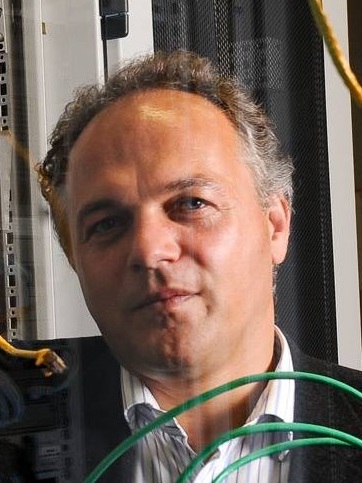}}]{Thomas C. Schmidt}
(Member, IEEE)  received the Ph.D. degree in mathematical physics from FU Berlin. He is a Professor of Computer Networks and Internet Technologies with the Hamburg University of Applied Sciences (HAW), where he heads the Internet Technologies Research Group. Prior to moving to Hamburg, he was a Director of a scientific computer centre in Berlin. Since then, he has continuously conducted numerous national and international research projects. He was the principal investigator in a number of EU, nationally funded and industrial projects as well as a Visiting Professor with the University of Reading, U.K. His continued interests lie in the development, measurement, and analysis of large-scale distributed systems like the Internet. He serves as co-editor and a technical expert in many occasions and he is actively involved in the work of IETF and IRTF. Together with his group, he pioneered work on an information-centric Industrial IoT and the emerging data-centric Web of Things. He is a co-founder of several large open source projects and a coordinator of the community developing the RIOT operating system—the friendly OS for the Internet of Things.
\end{IEEEbiography}

\begin{IEEEbiography}[{\includegraphics[width=1in,height=1.25in,clip,keepaspectratio]{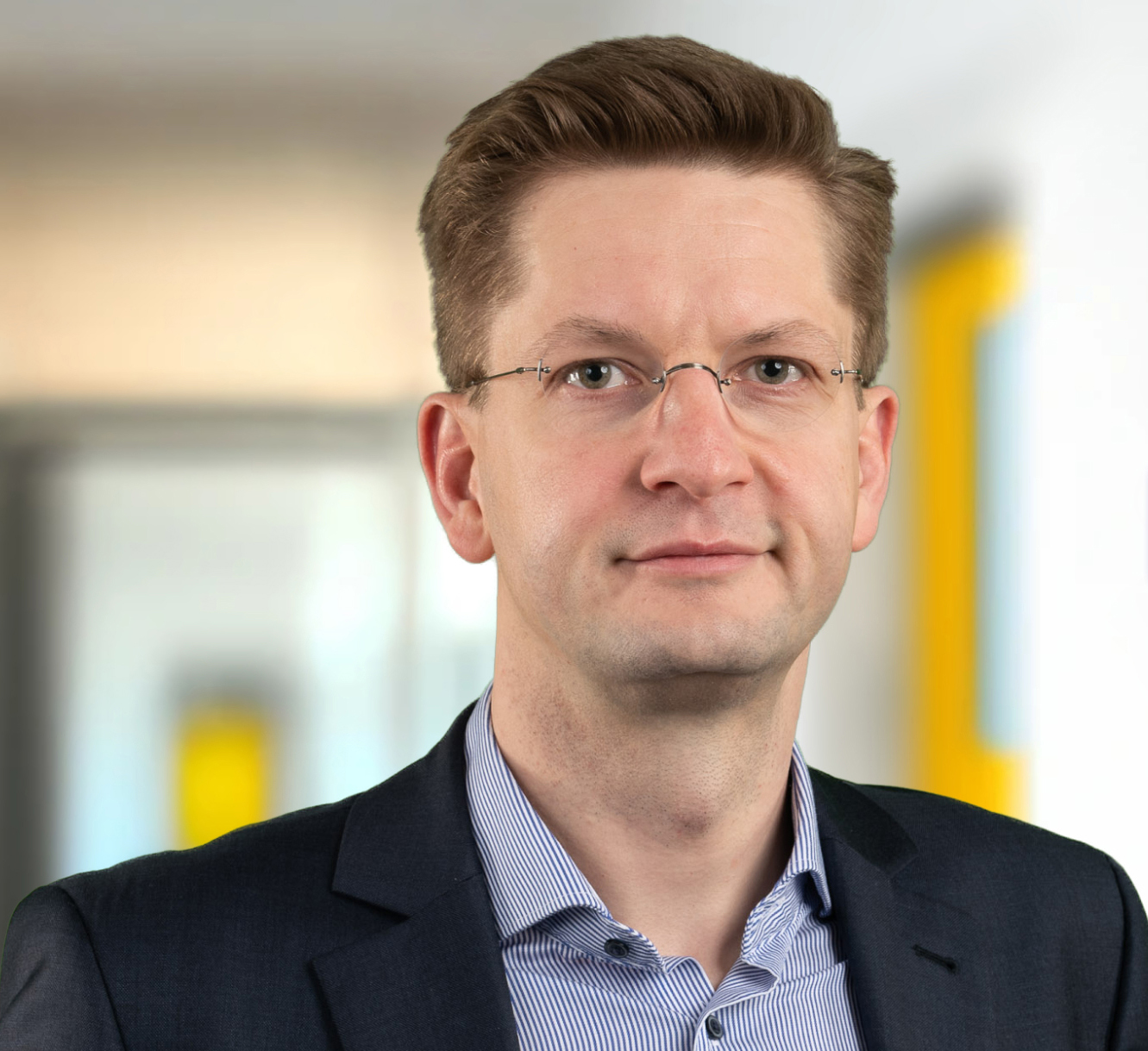}}]{Matthias W\"ahlisch}
(Member, IEEE) received the Ph.D. degree (Highest Hons.) in computer science from Freie Universität Berlin. He is a full professor and holds the Chair of Distributed and Networked Systems at the Faculty of Computer Science at TU~Dresden. He is also a Research Fellow of the Barkhausen Institut. His research and teaching focus on scalable, reliable, and secure Internet communication. This includes the design and evaluation of networking protocols and architectures, as well as Internet measurements and analysis. He is the PI of several national and international projects, supported by overall 8.2M EUR grant money. Since 2005, he has been active within IETF/IRTF. He published more than 160 peer-reviewed papers. He co-founded some successful open source projects, such as RIOT and RTRlib, where he is still responsible for the strategic development.
\end{IEEEbiography}

\end{document}